\documentclass[aps,pre,floatfix,longbibliography]{revtex4-2}
\usepackage{amsfonts}
\usepackage{amsmath}
\usepackage{amsthm}
\usepackage[x11names]{xcolor}
\usepackage{graphics}
\usepackage{graphicx,epstopdf}
\usepackage[export]{adjustbox}
\usepackage{subfigure}
\usepackage{amssymb}
\usepackage{lineno}
\usepackage[colorinlistoftodos]{todonotes}
\usepackage[T1]{fontenc}
\usepackage{amsmath,amssymb,epsfig,latexsym,graphicx,dcolumn}
\usepackage{natbib}

\usepackage{scalerel,stackengine}
\stackMath
\newcommand\reallywidehat[1]{%
	\savestack{\tmpbox}{\stretchto{%
			\scaleto{%
				\scalerel*[\widthof{\ensuremath{#1}}]{\kern-.6pt\bigwedge\kern-.6pt}%
				{\rule[-\textheight/2]{1ex}{\textheight}}
			}{\textheight}%
		}{0.5ex}}%
	\stackon[1pt]{#1}{\tmpbox}%
}
\begin{document}
	\setcounter{page}{1}
	\title{Stable, entropy-consistent, and localized artificial-diffusivity method\\ for capturing discontinuities}
	
	\author{Suhas S. Jain}\email{sjsuresh@stanford.edu}
	\affiliation{Center for Turbulence Research, Stanford University,  California, United States of America-94305}
	
	\author{Rahul Agrawal}
	\affiliation{Center for Turbulence Research, Stanford University,  California, United States of America-94305}
	
	\author{Parviz Moin}
	\affiliation{Center for Turbulence Research, Stanford University,  California, United States of America-94305}
	
	\date{\today}
	\begin{abstract}
		
		In this work, a localized artificial-viscosity/diffusivity method is proposed for accurately capturing discontinuities in compressible flows.
		There have been numerous efforts to improve the artificial diffusivity formulation in the last two decades, through appropriate localization of the artificial bulk viscosity for capturing shocks. However, for capturing contact discontinuities, either a density or internal energy variable is used as a detector. An issue with this sensor is that it not only detects contact discontinuities, but also falsely detects the regions of shocks and vortical motions. Using this detector to add artificial mass/thermal diffusivity for capturing contact discontinuities is hence unnecessarily dissipative. 
		To overcome this issue, we propose a sensor similar to the Ducros sensor (for shocks) to detect contact discontinuities, and further localize artificial mass/thermal diffusivity for capturing contact discontinuities. 
		
		The proposed method contains coefficients that are less sensitive to the choice of the flow problem. This is achieved by improved localization of the artificial diffusivity in the present method. A discretely consistent dissipative flux formulation is presented and is coupled with a robust low-dissipative scheme, which eliminates the need for filtering the solution variables. 
		The proposed method also does not require filtering for the discontinuity detector/sensor functions, which is typically done to smear out the artificial fluid properties and obtain stable solutions. Hence, the challenges associated with extending the filtering procedure for unstructured grids is eliminated, thereby, making the proposed method easily applicable for unstructured grids.
		Finally, a straightforward extension of the proposed method to two-phase flows is also presented.

	\end{abstract}

	\maketitle
	\textbf{Keywords}: compressible flows, turbulent flows, large-eddy simulation, artificial-viscosity method
	
	\section{Introduction} 
	
	Simulations of high-Mach-number compressible flows require a stable, and an accurate discontinuity-capturing method. As the Reynolds number increases, it is expected that the methods should capture both the turbulent structure near/far from the shock, and also capture the flow discontinuities such as shocks, contacts, and material interfaces, and also the interactions between the two.
	In incompressible flows, it is well known that low/non-dissipative schemes such as central schemes are required for accurate coarse-grained simulations such as the large-eddy simulation paradigm \cite{mittal1997suitability}. However, for capturing discontinuities, these standard central schemes introduce non-physical oscillations. Hence, the challenge associated with accurate simulations of compressible flows lies in the different needs for simulating turbulence and capturing discontinuities.

	Methods for capturing shocks/discontinuities can be broadly classified into three classes:
	(a) flux limiters and nonlinear schemes - these act at the scheme level, and therefore can be termed "implicit methods". 
	Both flux limiters \citep{shu1987tvb,wang2004spectral} and nonlinear schemes, such as essentially non-oscillatory (ENO) \citep{shu1988efficient,shu1989efficient}, weighted ENO (WENO) \citep{liu1994weighted,jiang1996efficient,cockburn1998essentially}, weighted nonlinear compact (WCN) \citep{deng2000developing,zhang2008development,nonomura2013robust} and targeted ENO (TENO) \citep{fu2016family} schemes, have been extensively used for capturing shocks in compressible flows.
	(b) artificial viscosity and artificial diffusivity - these act at the partial-differential equation (PDE) level, and therefore can be termed "explicit methods".
	(c) hybrid methods - they are a combination of implicit methods with central schemes that act in different regions of the domain.
	Some examples of hybrid approaches in the literature include the work of Ref. \cite{pirozzoli2011direct} where a modified Ducros sensor was used to switch between a central scheme and a WENO scheme, and Ref. \cite{hendrickson2018improved} where a modified Ducros sensor was used to switch between a central scheme and a modified Steger-Warming scheme.
	The comparison of various shock-capturing methods can be found in Ref. \cite{johnsen2010assessment,brehm2015comparison,lusher2019assessment}.

	In this work, a novel localized artificial-viscosity/diffusivity (AV/LAD)-based method for capturing shock and contact discontinuities is presented, and aimed towards capturing the aforementioned effects in compressible flows.
	Using an analogy between the Lax-Friedrichs (LF) flux and the artificial-viscosity methods, a discrete and consistent LF-type dissipative flux formulation is proposed for the LAD method. The proposed method satisfies the discrete kinetic energy-- and entropy-consistency conditions presented in Ref. \cite{jain2022kinetic}, and thus results in stable numerical simulations.
	We also propose new discontinuity detectors/sensors that localize where the artificial diffusivity is acting, and show that the proposed method is suitable for both direct numerical simulation (DNS) and large-eddy simulation (LES) of compressible turbulent flows with discontinuities.
	The sensors are designed in a way that the resulting method is less sensitive to the model coefficients (doesn't require tuning coefficients depending on the problem being solved, which is otherwise typically required for LAD methods). 
	Finally, the extension of the proposed method to compressible two-phase flows is also presented.
	
	In this work, we choose an artificial viscosity/diffusivity approach because of its advantages such as low cost, simplicity, and ease of implementation.
	Further, this method can be used with any underlying scheme, and more importantly, with central schemes which is beneficial for simulating turbulent flows. 
	Moreover, it also turns off naturally in smooth regions of the flow where there are no discontinuities.

	The artificial viscosity/diffusivity-based discontinuity-capturing methods can also be broadly categorized into two types of approaches. In one approach, artificial dissipation is directly added to all the equations without relating to the physical quantities. This is sometimes referred to as Laplacian viscosity. In the other approach, physical fluid properties, such as bulk viscosity, shear viscosity, and thermal conductivity, are augmented by artificial fluid properties to capture discontinuities. This is referred to as localized artificial viscosity/diffusivity (LAD). It is important to note that these two formulations can be expressed interchangeably. The major difference is the localization that is present in LAD, but typically absent in Laplacian viscosity approach.
	
	The idea behind artificial viscosity was first introduced by Ref. \cite{vonneumann1950method}. 
	Ref. \cite{jameson1981numerical} regularize solutions to the Euler equations. 
	To minimize the undesirable dissipation, Ref. \cite{tadmor1989convergence,kirby2006stabilisation} proposed a spectral vanishing viscosity (SVV) approach where only high-frequency components are damped.  
	Later, Ref. \cite{cook2004high} introduced the high-wavenumber viscosity approach, an idea similar to the SVV approach, but the dissipation was added in the physical space. 
	To further minimize the dissipation, they introduced different viscosities in the subsequent work \citep{cook2005hyperviscosity}, an artificial bulk viscosity (ABV) for shocks and artificial shear viscosity (ASV) for turbulence.
	Along similar lines, Ref. \cite{fiorina2007artificial} introduced artificial thermal diffusivity (ATD) for capturing temperature gradients and artificial species diffusivity for capturing species gradients using an entropy indicator function; and extended the LAD formulation for multicomponent reacting flows.
	Similarly, Ref. \cite{cook2007artificial} introduced artificial thermal conductivity for contact discontinuities and artificial diffusivity for material interfaces.
	More recently, the LAD formulation was also extended to curvilinear and anisotropic meshes by Ref. \cite{kawai2008localized}.
	
	The original LAD formulation used a strain rate-based indicator function in ABV to detect shocks. Ref. \cite{bhagatwala2008modified,mani2009suitability} replaced the strain rate-based indicator function with a negative dilatation-based indicator to localize ABV to shock regions and to turn it off in the regions of vortical motions.
	Ref. \cite{kawai2010assessment} augmented the ABV with a Ducros-type sensor to further localize the ABV for regions of shock and to turn it off in the regions of weak compression.
	
	Later, the LAD was coupled with a high-order flux reconstruction (FR) method \citep{miyaji2011compressible}, a spectral difference method \citep{premasuthan2014computation,premasuthan2014computation2}, and a discontinuous Galerkin method \citep{yu2013artificial} to extend the formulation to unstructured grids.
	Later, Ref. \cite{haga2019robust} adopted the FR+LAD framework and proposed a more sophisticated filter for robust simulations on unstructured grids.
	Ref. \cite{olson2013directional} proposed a LAD formulation where the artificial fluid properties are independently applied in each direction to avoid over-dissipation of discontinuities and numerical stiffness for high-aspect-ratio grids. 
	Ref. \cite{terashima2013consistent} proposed artificial mass diffusivity (AMD) as an alternative to ATD for multicomponent flows.
	The LAD has also been used with adaptive grid refinement, where the artificial diffusivity was used as an indicator for grid refinement \citep{nguyen2011adaptive,moro2016dilation}.
	More recently, Ref. \cite{jain2023assessment} further extended the LAD method by coupling it with a diffuse-interface method \citep{jain2020conservative} for the simulation of multiphase fluid flows and elastic-plastic deformation of solid-solid materials.

	It is worth mentioning the studies that used the other type of artificial (Laplacian) viscosity 
	\citep{jaffre1995convergence,bassi1997a,bassi1997high,baumann2000adaptive,hartmann2002adaptive,aliabadi2004alternative,hartmann2006adaptive,xin2006viscous}. 
	Ref. \cite{persson2006sub} used polynomial order dependent artificial viscosity for discontinuous Galerkin schemes.
	Ref. \cite{barter2007shock,reisner2013space} proposed an artificial viscosity based on a scalar PDE.
	Later, Ref. \cite{chandrashekar2013kinetic} proposed entropy-stable artificial viscosity, which has been extensively used in the recent works such as that of Ref. \cite{wintermeyer2018entropy}.
	Along similar lines, Ref. \cite{guermond2011entropy} proposed the entropy viscosity method, an artificial viscosity based on the local rate of generation of entropy. This was later extended to discontinuous finite element methods by Ref. \cite{zingan2013implementation} and discontinuous spectral element method by Ref. \cite{abbassi2014shock}.
	More recently, Ref. \cite{discacciati2020controlling} proposed an artificial neural network to predict the local artificial viscosity. 
	There are many more variants of this approach, but a drawback that is common to all these methods is that they use the same sensor for all the discontinuities, which could be overly dissipative for simulations of turbulent flows. Therefore, in this work, a LAD-based artificial viscosity is used because of the localized nature of the dissipation that is suitable for the simulations of turbulent flows.

	A characteristic of the LAD approaches is that they are most commonly used in conjunction with a high-order central scheme \citep{cook2005hyperviscosity,cook2007artificial,mani2009suitability, kawai2008localized, kawai2010assessment, lee2017localized, aslani2018localized,subramaniam2018high,adler2019strain,jain2023assessment}, with some exceptions \citep{haga2019robust}. However, in this work, we choose to use a second-order central-difference scheme with LAD since low-order central schemes are known to have some advantages for the simulation of turbulent flows \citep{moin2016suitability} due to their (a) non-dissipative nature, (b) low cost, (c) low aliasing error, (d) easy extension to unstructured grids, (e) ease of boundary treatment, and (f) improved stability. 
	In addition to these, the order of accuracy of high-order schemes can only be realized in an asymptotic regime, but the flow fields will be well resolved typically much before reaching this asymptotic regime.
	Moreover, for flows with discontinuities, the order of accuracy is also locally reduced around discontinuities.
	Recently, Ref. \cite{motheau2020investigation} compared a second-order Godunov method with a higher-order finite-volume WENO shock-capturing method and showed that it was significantly cost-effective to run a refined simulation with a second-order method than to run a coarse simulation using a higher-order method to obtain a solution of similar accuracy.


	\subsection{Challenges with the existing methods}
	The issues/challenges with the existing artificial viscosity methods are: 
	\begin{itemize}
		\item The LAD methods are generally used only with high-order numerical methods and have not been used with a low-order method; therefore, it is unclear how the LAD methods perform with a low-order method and there are no guidelines on how to choose the parameters when used with a low-order method.
		\item Some of the existing artificial viscosity formulations are too dissipative because of lack of proper localization, and hence are not suitable for the simulation of turbulent flows \citep{chandrashekar2013kinetic}, and others are less dissipative but do not add enough dissipation locally to resolve the jumps and result in inaccurate capturing of discontinuities.
		\item Some formulations are also not stable for high-Reynolds-number flows and require low-pass filters to eliminate oscillations, particularly with high-order numerical methods. Typically, these filters are used in the hope of achieving a stable method; however, the use of low-pass filtering might not necessarily always bring numerical stability to the method (see Ref. Ref. \cite{kawai2010assessment}). 
		\item In the existing LAD formulations, an artificial fluid property, $X^*$, is typically defined as
		\begin{equation*}
			X^* \sim C_X \Delta^{r+2} \overline{|\nabla^r s_X|}f_X,
		\end{equation*}
		where $C_X$ is a model constant, $\Delta$ is the local grid size, $s_X$ is a discontinuity indicator function with an additional localization sensor $f_X$. The overbar denotes a Gaussian filtering operation which is required to obtain a smooth artificial fluid property (particularly for large values of $r$) to achieve a stable method \citep{abbassi2014shock}. However, it is not trivial to extend this filtering operation to unstructured grids for simulations in complex geometries \citep{haga2019robust}.
		\item In the existing formulations, the artificial thermal/mass diffusivity (ATD/AMD) added to capture contact discontinuities is also active in the regions of shocks and vortical motions due to the choice of indicator function and lack of proper localization. This adds unnecessary additional dissipation by ATD/AMD in the regions of shock where ABV is already acting to capture shocks and in the regions of unresolved eddies where a subgrid model is already active.
		\item Additionally, the coefficients $C_X$ may also require problem-dependent tuning \citep{subramaniam2018high,jain2023assessment}, which is primarily due to the insufficient localization of the added artificial dissipation. For example, the coefficient values that are tuned for the simulation of turbulent flows with weak shocks/shocklets might not be appropriate for the simulation of stronger shocks, and similarly, the coefficient values that are tuned to capture stronger shocks could be too dissipative for the simulation of turbulent flows. 
		\item It is also common to turn on and off the artificial fluid properties \citep{haga2019robust} depending on the problem being solved, which is not predictive.
	\end{itemize}
	The proposed method in this work aims to address these challenges with existing AV/LAD methods. 
	
	A preliminary version of this work has been published as a technical report in the annual publication of the Center for Turbulence Research \citep{jain2021stable}.
	The rest of this paper is organized as follows. Section \ref{sec:pde_model} contains the proposed localized artificial-viscosity/diffusivity model along with the details of the sensors used, a new sensor for detecting contact discontinuities, the coefficient values used, and an extension to two-phase flows. Section \ref{sec:discretization} contains the consistency conditions for kinetic energy\textendash and entropy-consistent discretization and a consistent dissipative flux formulation. Section \ref{sec:results} contains the simulation results using the proposed method, and finally the concluding remarks are presented in Section \ref{sec:conclusions}.

	
	

	
	

	\section{Proposed artificial-viscosity/diffusivity method\label{sec:pde_model}}

	
	The idea behind a LAD method is to augment the physical fluid properties with the grid-dependent artificial fluid properties locally in the regions of the flow where discontinuities such as shocks (artificial bulk viscosity), contacts (artificial mass/thermal diffusivity), and eddies (artificial shear viscosity) are not resolved by the grid \citep{cook2005hyperviscosity,cook2007artificial}. 
	
	For capturing shocks on the grid, an artificial bulk viscosity (ABV), $\beta^*$, is appended to the physical bulk viscosity, $\beta_p$, as
	\begin{equation}
		\beta = \beta_p + \beta^*.
		\label{eq:ABV}
	\end{equation}
	Initially, a strain-rate-based sensor was used to detect shocks \citep{cook2007artificial}, but this made the ABV active in regions away from shocks and in turbulent flows.  
	Since shocks are associated with high values of negative dilatation, to reduce the ABV away from shocks, Ref. \cite{bhagatwala2008modified} and Ref. \cite{mani2009suitability} proposed a dilatation-based sensor. Further, Ref. \cite{kawai2010assessment} appended the Ducros sensor \citep{ducros1999large} to ABV to reduce ABV in the regions of high enstrophy that represent turbulent motions and to localize ABV further. 
	For unresolved vortical motions, an artificial shear viscosity (ASV), $\mu^*$, is used as a subgrid model where it is appended to the physical shear viscosity, $\mu_p$, as
	\begin{equation}
		\mu = \mu_p + \mu^*.
	\end{equation}
	Typically, the magnitude of the strain rate tensor is used as an indicator function \citep{cook2007artificial}. Alternatively, an eddy-viscosity model \citep{smagorinsky1963general,germano1991dynamic} can be used as a subgrid-scale model for unresolved eddies, which is a more widely adopted approach, and we use this approach in this work. This is motivated by the observations that a dynamic subgrid-scale model appropriately turns on/off its activity based on the amount of resolved turbulence, whereas a numerical dissipation approach does not account for it appropriately. 
	Similarly, for capturing unresolved contact discontinuities, an artificial thermal diffusivity (ATD), $\kappa^*$, is appended to the physical thermal diffusivity, $\kappa_p$, \citep{cook2007artificial,kawai2010assessment,jain2023assessment} as
	\begin{equation}
		\kappa = \kappa_p + \kappa^*.
	\end{equation}
	Alternatively, artificial mass diffusivity (AMD), $D^*$, can be used in the place of ATD  \citep{terashima2013consistent} where artificial terms such as $\vec{\nabla}\cdot(D^* \vec{\nabla}\rho)$, $\vec{\nabla}\cdot(D^* \vec{\nabla}\rho \otimes \vec{u})$, and $\vec{\nabla}\cdot\{(|\vec{u}|^2/2)D^* \vec{\nabla}\rho\}$ are consistently added to the mass, momentum, and energy equations.  These consistent corrections for the momentum and energy equations are similar to the consistent corrections used in capturing material interfaces in two-phase flows \citep{jain2020conservative}.
	We adopt the AMD approach over ATD because it can be shown that AMD satisfies the interface equilibrium condition (IEC) or sometimes also referred to as the pressure equilibrium, an important thermodynamic consistency condition for robust numerical simulations of compressible two-phase flows \citep{jain2020conservative}, with the five-equation model which the ATD does not satisfy (see Section \ref{sec:two_phase}). However, for single-phase flows, both ATD and AMD are equally applicable and will yield similar results. 
	
	\subsection{Proposed model}
	
	The system of conservation equations (mass, momentum, and energy) along with the proposed artificial-viscosity method can be written as
	\begin{equation}
		\frac{\partial \rho}{\partial t} + \frac{\partial \rho u_j}{\partial x_j} = A_{\rho},
		\label{eq:masss}
	\end{equation}
	\begin{equation}
		\frac{\partial \rho u_i}{\partial t} + \frac{\partial \rho u_i u_j}{\partial x_j} + \frac{\partial p}{\partial x_i} = \frac{\partial \tau_{ij}}{\partial x_j} + \rho g_i + A_{\rho u},    
		\label{eq:moms}
	\end{equation}
	and
	\begin{equation}
		\begin{aligned}
			\frac{\partial E}{\partial t} + \frac{\partial \left(E + p \right) u_j}{\partial x_j} = \frac{\partial \tau_{ij} u_i}{\partial x_j} + \rho u_i g_i + A_{E}, 
			\label{eq:energys}
		\end{aligned}
	\end{equation}
	where $\rho$ is the density, $p$ is the pressure, $u$ is the velocity, $E=\rho(e + u_i u_i/2)$ is the total energy and $e$ is the internal energy, $\tau_{ij}$ is the stress tensor, and $g_i$ represents a generic body force. Throughout this paper, $i$ and $j$ represent Einstein indices, and $x$ and $t$ represent space and time coordinates, respectively. In Eqs. \eqref{eq:masss}-\eqref{eq:energys}, $A_{\rho}$, $A_{\rho u}$, and $A_E$ are the artificial terms added to the mass, momentum, and energy equations, respectively, to capture shocks and contact discontinuities. They can be written as 
	\begin{equation}
		A_{\rho} = {\frac{\partial}{\partial x_j} \left(D^* \frac{\partial \rho}{\partial x_j} \right)},
		\label{eq:art_mass}
	\end{equation}
	\begin{equation}
		A_{\rho u} = {\frac{\partial}{\partial x_j} \left(D^* u_i \frac{\partial \rho}{\partial x_j} \right)}    + {\frac{\partial}{\partial x_j} \left(\beta^* \frac{\partial u_k}{\partial x_k} \delta_{ij} \right)},
		\label{eq:art_mom}
	\end{equation}
	and
	\begin{equation}
		\begin{aligned}
			A_{E} = {\frac{\partial}{\partial x_j} \left[D^* \frac{\partial \rho}{\partial x_j} \left(\frac{u_k u_k}{2}\right)\right]} 
			+ {\frac{\partial}{\partial x_j} \left(D^* \frac{\partial \rho e}{\partial x_j} \right)} 
			+ {\frac{\partial}{\partial x_j} \left(\beta^* \frac{\partial u_k}{\partial x_k} \delta_{ij} u_i \right)}.
			\label{eq:art_energy}
		\end{aligned}
	\end{equation}
	Here, Eq. \eqref{eq:art_mass}, the first term in Eq. \eqref{eq:art_mom}, and the first two terms in Eq. \eqref{eq:art_energy} are 
	added to capture contact discontinuities; and the second term in Eq. \eqref{eq:art_mom} and the last term in Eq. \eqref{eq:art_energy} are added to capture shocks. 
	These consistent terms in the momentum and energy equations can be derived similarly to the derivation of interface-regularization terms as described in Ref. \cite{jain2020conservative}. 
	The consistent terms in the momentum and energy equations are introduced such that there is no spurious contribution to the total kinetic energy of the system.
	Note that, in this work, the ASV is not used (or is equivalently set to zero). Instead, a dynamic Smagorinsky model is used for unresolved eddies, and in Section \ref{sec:sgs_results}, the sensitivity of the results to the presence of a dynamic subgrid-scale model is investigated. 
	
	\subsection{Dynamic subgrid scale model \label{sec:dsm}}
	
	For the sake of completeness, the details of the subgrid-scale model are provided here. The resolved turbulent stresses, $L_{ij} = - \widehat{\overline{u}_i\overline{u}_j} + \widehat{\overline{u}_i} \; \widehat{\overline{u}_j}$ [$\widehat{(\cdot)}$ denotes test-filter operation], are related to the modeled stresses, $M_{ij}$, in the ``test window'' \citep{germano1991dynamic} as
	\begin{equation}
		L_{ij} = 2\left(C_s \Delta \right)^2 \left( {\frac{\widehat{\Delta}^2}{\Delta^2}} \widehat{|S|}\widehat{S_{ij}} - \widehat{|S|S_{ij}}\right) = 2\left(C_s \Delta \right)^2 M_{ij},
		\label{eqn:dsm}
	\end{equation}
	where $\widehat{\Delta}$ and $\Delta$ denote test-level and grid-level filter widths, respectively. $C_s$ is subgrid-scale the model coefficient, and $S_{ij}$ is the rate of strain tensor from resolved LES fields, and $|S|$ is its magnitude.   Ref. \cite{lilly1992proposed} proposed a least-squares solution of this system, leading to the expression for the model coefficient
	\begin{equation}
		(C_s \Delta)^2 =  \frac{L_{ij} M_{ij}}{2 M_{ij} M_{ij}}.
	\end{equation}
	In this work, both the numerator and denominator have been averaged in the volume (since for a periodic box all three spatial directions are homogeneous), to arrive at 
	\begin{equation}
		(C_s \Delta)^2 =  \frac{\langle L_{ij} M_{ij}\rangle}{\langle 2 M_{ij} M_{ij}\rangle}, 
	\end{equation}
	where $\langle \cdot \rangle$ is the volumetric-averaging operator. The final form of the eddy-viscosity model is then written as, 
	\begin{equation}
		\mu^{*} = (C_s \Delta)^2 |S| 
	\end{equation}

	
	
	
	\subsection{Artificial fluid properties}
	
	The artificial diffusivities used in this work can be defined as 
	\begin{equation}
		\begin{aligned}
			D^* = C_{D}  {\frac{1}{\rho} \left|\sum_j \frac{\partial^r \rho}{\partial x^r_j} (\Delta x_j)^{r+1}\right|}
			\left(|u_j| + c_s\right)
			f_{D},
		\end{aligned}
		\label{eq:mass_diff}
	\end{equation}
	and
	\begin{equation}
		\beta^* = C_{\beta} {\rho  \left|\sum_j\frac{\partial^r \theta}{\partial x^r_j} (\Delta x_j)^{r+2}\right| H(-\theta)} f_{\beta},
		\label{eq:bulk_diff}
	\end{equation}
	where $\Delta x$ is the grid size; $C_D$ and $C_{\beta}$ are the model coefficients for AMD and ABV, respectively; $c_s$ is the speed of sound; $\theta=\vec{\nabla}\cdot \vec{u}$ is the dilatation; and $f_D$ and $f_{\beta}$ are the localization sensors. Here, $\rho$ and $\theta$ act as indicator functions in AMD and ABV to detect contact discontinuities and shocks, respectively. The Heaviside function $H(-\theta)$ turns off the ABV in the regions of expansion. It is important to note that the Gaussian filtering operation is not performed on these artificial fluid properties, unlike the previous LAD methods, where the Gaussian filtering was required to obtain stable solutions. This makes it easy for the present method to be extended for unstructured grids.

	The localization sensor in ABV is given as
	\begin{equation}
		f_{\beta} = \left(\frac{\theta^2}{\theta^2 + a\omega_i\omega_i + \varepsilon}\right),
		\label{eq:Ducros_sensor}
	\end{equation}
	which is a modified version ($a>1$) of the Ducros sensor \citep{ducros1999large}, where $\varepsilon=1e^{-15}$ is a small number added to prevent division by zero. The original Ducros sensor ($a=1$) was used by Ref. \cite{kawai2010assessment} to localize ABV for capturing shocks. The idea behind using this sensor is to identify the regions of weak compression where the enstrophy could be non-negligible compared to the local dilatation, and to turn off ABV in those regions. The modification ($a>1$) here only makes the localization sensor stronger.
	
	\subsection{A sensor for detecting contact discontinuities \label{sec:contact_sensor}}
	
	In all the previous studies, either density or internal energy is used as an indicator function in AMD/ATD to detect contact discontinuities \citep{cook2007artificial,kawai2008localized,bhagatwala2008modified,kawai2010assessment,miyaji2011compressible,yu2013artificial,olson2013directional,terashima2013consistent,premasuthan2014computation,premasuthan2014computation2,lee2017localized,subramaniam2018high,haga2019robust}. 
	An issue with this approach is that this indicator will not only detect contact discontinuities, but will also detect shocks and vortical motions where there are jumps in density and internal energy (finite gradients). Therefore, to overcome this issue, we propose a sensor\textemdash motivated by the idea behind the Ducros sensor\textemdash for detecting contact discontinuities as
	\begin{equation}
		f_{D} =  \left[\frac{\lvert\frac{\partial \rho}{\partial x_j}\rvert^2}{\lvert\frac{\partial \rho}{\partial x_j}\rvert^2 + a(\theta^2 + \omega_i\omega_i)\left(\frac{\rho}{|\vec{u}|}\right)^2 + \varepsilon}\right].
		\label{eq:new_sensor}
	\end{equation}
	This sensor by construction can detect contact discontinuities very effectively and can distinguish it from the regions of shocks (high dilatation) and vortical motions (high enstrophy). This sensor will turn on only when there is a jump in density, and will turn off in the regions of high dilatation or enstrophy.  
	In this work, the purpose of adding this new sensor $f_D$ is to further localize AMD by making it active only in the regions containing contact discontinuities and to turn off AMD in the regions with high enstrophy and high dilatation, which represent vortical motions and shocks, respectively. Note that the proposed $f_D$ can also be used with an ATD approach for capturing contact discontinuities without the loss of generality.

	\subsection{Problem-independent coefficients and the choice of $r$ \label{sec:coefficients}}

	The coefficients $C_D$ and $C_{\beta}$ are generally dependent on the numerical scheme used. In this work, for the second-order central schemes, we use the values $C_{D}\approx0.5$, $C_{\beta}\approx100$, and $a\approx100$. For the flows considered, these values are shown to produce fairly similar results, as long as the values are changed by less than an order of magnitude (see Sections \ref{sec:sensitivity1} and \ref{sec:sensitivity2}). This is primarily due to the use of sophisticated sensors in Eqs. \eqref{eq:new_sensor} and \eqref{eq:Ducros_sensor} that are responsible for adding dissipation locally only in the regions where it is needed the most. Hence, the values of these coefficients $C_D$ and $C_{\beta}$ need to be determined only once for a particular scheme. Without the use of the new sensor in Section \ref{sec:contact_sensor}, the current choice of values for $C_{D}$ and $C_{\beta}$ is too dissipative for turbulent flows with weak compressibility (see Figure \ref{fig:decay_HIT} and the discussion in Section \ref{sec:decaying_HIT}), and if the values for these coefficients are reduced to make them suitable for turbulent flows, the dissipation is insufficient to capture stronger shocks. However, in the presence of the proposed sensor, the current choice of model coefficients can be used without the need for tuning it for each problem, e.g., stronger shocks and turbulent flows with weak compressibility. 
	
	Furthermore, in this study, a value of $r=1$ is used. Note that a higher value of $r$ was used in other studies that use higher-order schemes \citep{cook2007artificial,mani2009suitability, kawai2008localized, kawai2010assessment, lee2017localized, aslani2018localized,subramaniam2018high,adler2019strain,jain2023assessment}. In previous studies, a higher value of $r$ was suggested because it results in dissipating only the higher-wavenumber content that is not resolved by the scheme \citep{cook2004high}. However, in this study, with the use of a low-order scheme, the motivation to use $r=1$ (a lower value of $r$) is to have a sensor that is more localized only at the discontinuities and not to dissipate unresolved scales, unlike the motivation for higher-order schemes. For lower-order schemes, a higher value of $r>2$ makes the sensor active in other regions in the domain where there is no shock or a contact discontinuity, and this would not be ideal for the simulation of turbulent flows. 

	\subsection{Extension to two-phase flows \label{sec:two_phase}} 
	
	In this section, the proposed artificial-viscosity method is extended for two-phase flows. Recently, a conservative version of the diffuse-interface model that can be discretized using a central-differencing scheme was proposed to simulate compressible two-phase flows \citep{jain2020conservative,jain2022accurate,jain2023assessment}. A five-equation model based on  Ref. \cite{allaire2002five} and Ref. \cite{kapila2001two} was proposed in Ref. \cite{jain2020conservative} and used with a low-order central scheme, and a four-equation model was proposed in Ref. \cite{jain2023assessment} and used with a high-order central scheme. 
	
	Here, we present an extension of the proposed artificial-diffusivity method for two-phase flows. The system of conservation equations for volume fraction, mass of each phase, momentum, and total energy can be written as
	\begin{equation}
		\frac{\partial \phi_1}{\partial t} + \frac{\partial u_j \phi_1}{\partial x_j} = (\phi_1 + \zeta_1)\frac{\partial u_j}{\partial x_j} + \frac{\partial a_{1j}}{\partial x_j} + {\frac{\partial}{\partial x_j} \left(D^* \frac{\partial \phi_l}{\partial x_j} \right)}, 
		\label{eq:volumef}    
	\end{equation}
	\begin{equation}
		\frac{\partial \rho_l\phi_l}{\partial t} + \frac{\partial u_j \rho_l \phi_l}{\partial x_j} = \frac{\partial R_{lj}}{\partial x_j} + {\frac{\partial}{\partial x_j} \left(D^* \frac{\partial \rho_l \phi_l}{\partial x_j} \right)},  \hspace{0.5cm} l=1,2,
		\label{eq:massf}
	\end{equation}
	\begin{equation}
		\begin{aligned}
			\frac{\partial \rho u_i}{\partial t} + \frac{\partial \rho u_i u_j}{\partial x_j} + \frac{\partial p}{\partial x_i} = \frac{\partial u_i f_j}{\partial x_j} + \frac{\partial \tau_{ij}}{\partial x_j} + \sigma \kappa \frac{\partial \phi_1}{\partial x_i} + \rho g_i\\
			+{\frac{\partial}{\partial x_j} \left(D^* u_i \frac{\partial \rho}{\partial x_j} \right)}
			+ {\frac{\partial}{\partial x_j} \left(\beta^* \frac{\partial u_k}{\partial x_k} \delta_{ij} \right)},
		\end{aligned}
		\label{eq:momf}
	\end{equation}
	\begin{equation}
		\begin{aligned}
			\frac{\partial E}{\partial t} + \frac{\partial \left(E + p \right) u_j}{\partial x_j} = \frac{\partial \tau_{ij} u_i}{\partial x_j} + \frac{\partial k f_j}{\partial x_j} + \sum_{l=1}^2 \frac{\partial \rho_l h_l a_{lj}}{\partial x_j} + \sigma \kappa u_i\frac{\partial \phi_1}{\partial x_i} + \rho u_i g_i\\
			+ {\frac{\partial}{\partial x_j} \left[D^* \frac{\partial \rho}{\partial x_j} \left(\frac{u_k u_k}{2}\right)\right]}
			+ {\frac{\partial}{\partial x_j} \left(D^* \frac{\partial \rho e}{\partial x_j} \right)}
			+ {\frac{\partial}{\partial x_j} \left(\beta^* \frac{\partial u_k}{\partial x_k} \delta_{ij} u_i \right)},
		\end{aligned}
		\label{eq:energyf}
	\end{equation}
	where $\phi_l$ is the volume fraction of phase $l$ that satisfies the condition $\sum_{l=1}^2 \phi_l=1$; $\rho_l$ is the density of phase $l$; $\rho$ is the total density, defined as $\rho=\sum_{l=1}^2\rho_l\phi_l$; $\vec{u}$ is the velocity; $p$ is the pressure; $e$ is the specific mixture internal energy, which can be related to the specific internal energy of phase $l$, $e_l$, as $\rho e=\sum_{l=1}^2 \rho_le_l$; $k=u_iu_i/2$ is the specific kinetic energy; $E=\rho(e+k)$ is the total energy of the mixture per unit volume; and the function $\zeta_1$ is given by
	\begin{equation}
		\zeta_1 = \frac{\rho_2 c_2^2 - \rho_1 c_1^2}{\frac{\rho_1 c_1^2}{\phi_1} + \frac{\rho_2 c_2^2}{\phi_2}},
	\end{equation}
	for the Kapila's five-equation model and is $\zeta_1=0$ in the Allaire's five-equation model, where $c_l$ is the speed of sound for phase $l$. 
	In Eq. (\ref{eq:energyf}), $h_l=e_l+p/\rho_l$ represents the specific enthalpy of phase $l$ and can be expressed in terms of $\rho_l$ and $p$ using the stiffened-gas equation of state as
	\begin{equation}
		h_l=\frac{(p + \pi_l)\gamma_l}{\rho_l(\gamma_l - 1)}.
		\label{eq:enthalpy}
	\end{equation}
	In Eqs. (\ref{eq:volumef})-(\ref{eq:energyf}), $\sigma$ is the surface-tension coefficient, $\kappa=-\vec{\nabla}\cdot\vec{n}_1$ is the curvature of the interface, $\vec{n}_l$ is the normal vector of the interface for phase $l$, $\vec{g}$ is the gravitational acceleration, and $\vec{a}_l$ is the volumetric interface-regularization flux for phase $l$ which is responsible for keeping the finite thickness of the material interface, and this satisfies the condition $\vec{a}(\phi_1)=-\vec{a}(\phi_2)$. A conservative phase-field model or an accurate conservative phase-field models can be used as the interface-regularization fluxes, as was proposed in Ref. \cite{jain2020conservative} and Ref. \cite{jain2022accurate}, respectively. 
	$\vec{R}_l=\rho_l\vec{a}_l$ is the consistent regularization flux for the mass of phase $l$,  
	and $\vec{f}=\sum_{l=1}^2 \vec{R}_l=\sum_{l=1}^2 \rho_l\vec{a}_l$ is the net consistent regularization flux for the mixture mass.

	The ABV in Eq. \eqref{eq:bulk_diff} requires no modification for two-phase flows, but AMD requires modification because of the jump in density at the material interface which will activate the AMD. Since the diffuse-interface model is already active to capture material interface, AMD has to be turned off at the interface. This can be easily achieved by replacing the gradients of density in Eqs. \eqref{eq:art_mass}, \eqref{eq:new_sensor} with a volume weighted gradients of phase density as
	\begin{equation*}
		\frac{\partial \rho}{\partial x_j} \rightarrow \sum_l \phi_l \frac{\partial \rho_l}{\partial x_j},
	\end{equation*}
	which will prevent AMD from activating at material interfaces.

	\section{Discrete kinetic energy-- and entropy-consistent dissipative flux formulation\label{sec:discretization}}
	
	We first show an analogy between the Lax-Friedrichs (LF) flux and the artificial-viscosity methods, and then use this idea to develop a kinetic energy--and entropy-consistent flux formulation for the proposed artificial-viscosity method. Consider a generic conservation equation of the form
	\begin{equation}
		\frac{\partial c}{\partial t} + \frac{\partial f(c)}{\partial x} = 0,
		\label{eq:gen_cons}
	\end{equation}
	where $c$ is a conserved quantity, and $f(c)$ is the flux function. Using the explicit Euler (EE) time-advancement scheme and the second-order central scheme, the discrete form of the equation can be written as
	\begin{equation*}
		c^{n+1}_m = c^{n}_m - \frac{\Delta t}{2 \Delta x} \left[ f(c^{n}_{m+1}) - f(c^{n}_{m-1}) \right],
	\end{equation*} 
	where $m$ is the grid index.
	Now, replacing $c^{n}_m$ with $(c^{n}_{m+1} + c^{n}_{m})/2$, and rewriting in conservation (flux) form, we obtain
	\begin{equation*}
		c^{n+1}_m = c^{n}_m - \lambda \left( \hat{f}_{m+\frac{1}{2}} - \hat{f}_{m-\frac{1}{2}} \right),
	\end{equation*}
	where the numerical flux, $\hat{f}_{m+{1}/{2}}$, is the well known LF flux \citep{leveque1992numerical}, given by
	\begin{equation}
		\hat{f}_{m+\frac{1}{2}} = \left[ \frac{f(c^{n}_{m+1}) + f(c^{n}_{m-1})}{2}\right] - \frac{1}{2\lambda} (c^{n}_{m+1} - c^{n}_{m}), 
	\end{equation}
	and $\lambda = \Delta t/\Delta x$ has units of inverse velocity. Now consider the same conservation equation in Eq. \eqref{eq:gen_cons} augmented with a generic artificial-viscosity fluid property, $\epsilon^*$, as
	\begin{equation*}
		\frac{\partial c}{\partial t} + \frac{\partial f(c)}{\partial x} = \frac{\partial}{\partial x} \left[\epsilon^* \frac{\partial c}{\partial x} \right].
	\end{equation*}
	Using EE and second-order central schemes and writing in conservation form, we arrive at
	\begin{equation*}
		c^{n+1}_m = c^{n}_m - \lambda \left[ \hat{f}_{m+\frac{1}{2}} - \hat{f}_{m-\frac{1}{2}} \right],
	\end{equation*}
	where the numerical flux is
	\begin{equation}
		\hat{f}_{m+\frac{1}{2}} = \left[ \frac{f(c^{n}_{m+1}) + f(c^{n}_{m-1})}{2}\right] - \frac{\epsilon^*_{m+\frac{1}{2}}}{\Delta x} (c^{n}_{m+1} - c^{n}_{m}).
	\end{equation}
	Note that if $\epsilon^* = 2 \Delta x/\lambda$, then the LF flux and the artificial-viscosity method are identical (provided that a second-order central scheme is used for the discretization of the artificial-viscosity terms). 
	
	The LF flux is known to be entropy stable, but it is also highly dissipative. Therefore, the idea proposed in this work is to replace the non-dissipative central-flux in the LF flux with a robust kinetic energy--and entropy-preserving (KEEP) type flux in Ref. \cite{jain2022kinetic} and to further localize the dissipative part of the LF flux, only to those regions where they are needed, with the use of sensors. The new proposed flux can then be represented as
	\begin{equation}
		\hat{f}_{m+\frac{1}{2}} = \left.\hat{f}_{m+\frac{1}{2}}\right\rvert_{KEEP}  - \hat{A}_{m+\frac{1}{2}}^d,
	\end{equation}
	where $\hat{A}_{m+{1}/{2}}^d$ is the discrete dissipative flux for the artificial term, given by
	\begin{equation}
		\hat{A}_{m+\frac{1}{2}}^d = \frac{\epsilon^*_{m+\frac{1}{2}}}{\Delta x} \left(c^{n+1}_{m+1} - c^{n}_{m}\right). 
		\label{eq:diss_flux}
	\end{equation}
	Here, $\epsilon^*_{m+{1}/{2}}$ represents a localized artificial-fluid property. In this work, the non-dissipative central flux is replaced with the second-order KEEP scheme of Ref. \cite{jain2022kinetic}. However, in general, this central flux can be replaced with any other robust low/non-dissipative flux \citep{chandrashekar2013kinetic,kennedy2008reduced,kuya2018kinetic,coppola2019numerically}.

	Ref. \cite{chandrashekar2013kinetic} explored a similar idea by replacing the non-dissipative central flux with a kinetic energy-preserving (KEP) flux. But the dissipative flux in their approach was not localized and was present everywhere in the domain. Another difference is that they used a scalar dissipation of momentum to capture all discontinuities. 
	The scalar dissipation of momentum not only acts on the dilatational motion at the shocks, but also dissipates the vortical structures, which makes the method even more dissipative and unsuitable for the simulation of turbulent flows. They concluded that this approach was too dissipative for turbulent flows. 
	


	\subsection{Discrete consistency conditions}
	
	The discrete consistency conditions between the mass, momentum, kinetic energy, and internal energy convective fluxes and artificial fluxes (for capturing material interfaces) were proposed by Ref. \cite{jain2020keep}. Following a similar procedure, the consistency conditions can be further extended to include dissipative fluxes, which represent the artificial fluid diffusivity, in this work. If the full flux in the mass equation [Eq. \eqref{eq:masss}] is written as
	\begin{equation}
		\hat{C}^f_j\rvert_{(m\pm\frac{1}{2})} = \hat{C}_j\rvert_{(m\pm\frac{1}{2})} + \hat{C}_j^{'}\rvert_{(m\pm\frac{1}{2})},
		\label{eq:mass_cons}
	\end{equation}
	where $\hat{C}_j\rvert_{(m\pm{1}/{2})}$ is the convective part and $\hat{C}_j^{'}\rvert_{(m\pm{1}/{2})}$ is the dissipative AMD contribution, then, the momentum- and kinetic energy--consistency conditions are given by
	\begin{equation}
		\hat{M}^f_{ij}\rvert_{(m\pm\frac{1}{2})} = \left(\hat{C}_j\rvert_{(m\pm\frac{1}{2})} + \hat{C}_j^{'}\rvert_{(m\pm\frac{1}{2})} \right) \overline{u}_{i}^{(m\pm\frac{1}{2})} + \hat{M}^{'}_{ij}\rvert_{(m\pm\frac{1}{2})},
		\label{eq:mom_cons}
	\end{equation}
	and
	\begin{equation}
		\hat{K}^f_j\rvert_{(m\pm\frac{1}{2})} = \left(\hat{C}_j\rvert_{(m\pm\frac{1}{2})} + \hat{C}_j^{'}\rvert_{(m\pm\frac{1}{2})} \right) \frac{u_i\rvert_{(m\pm1)} u_i\rvert_{(m)}}{2} + \overline{u}_{i}^{(m\pm\frac{1}{2})} \hat{M}^{'}_{ij}\rvert_{(m\pm\frac{1}{2})},
		\label{eq:ke_cons}
	\end{equation}
	where $\hat{M}^{'}_{ij}\rvert_{(m\pm{1}/{2})}$ represents the additional momentum dissipative flux (ABV contribution) and the overbar $\overline{\cdot}^{(m\pm{1}/{2})}$ denotes an arithmetic average of a quantity at $m$ and $m\pm1$;
	%
	
	\subsection{Discrete fluxes for the proposed method}
	
	Using the consistency conditions in Eqs. \eqref{eq:mass_cons}-\eqref{eq:ke_cons}, and following the notation used in Eq. \eqref{eq:diss_flux} for the LF-type dissipative fluxes, the proposed artificial-viscosity method can be written in discrete flux form as
	\begin{gather}
		\hat{A_{\rho}}^{d}_j\rvert_{(m\pm\frac{1}{2})} =  \hat{C}_j^{'}\rvert_{(m\pm\frac{1}{2})} = - \frac{D^*_{m\pm\frac{1}{2}}}{\Delta x_j} (\Delta_j \rho), \\
		\hat{A_{\rho u}}^d_{ij}\rvert_{(m\pm\frac{1}{2})} = \hat{C}_j^{'}\rvert_{(m\pm\frac{1}{2})} \overline{u}_{i}^{(m\pm\frac{1}{2})} + \hat{M}^{'}_{ij}\rvert_{(m\pm\frac{1}{2})} 
		=\left( - \frac{D^*_{m\pm\frac{1}{2}}}{\Delta x_j} (\Delta_j \rho)  \right) \overline{u}_{i}^{(m\pm\frac{1}{2})} - \beta^*_{m\pm\frac{1}{2}} \reallywidehat{\frac{\partial u_k}{\partial x_k}}\rvert_{(m\pm\frac{1}{2})} \delta_{ij},\\
		\begin{split}
			\hat{K}^d_j\rvert_{(m\pm\frac{1}{2})} = 
			\hat{C}_j^{'}\rvert_{(m\pm\frac{1}{2})} \frac{u_i\rvert_{(m\pm1)} u_i\rvert_{(m)}}{2} + \overline{u}_{i}^{(m\pm\frac{1}{2})} \hat{M}^{'}_{ij}\rvert_{(m\pm\frac{1}{2})} \\
			=\left(- \frac{D^*_{m\pm\frac{1}{2}}}{\Delta x_j} (\Delta_j \rho)  \right) \frac{u_i\rvert_{(m\pm1)} u_i\rvert_{(m)}}{2} + \left( - \beta^*_{m\pm\frac{1}{2}} \reallywidehat{\frac{\partial u_k}{\partial x_k}}\rvert_{(m\pm\frac{1}{2})} \delta_{ij}\right) \overline{u}_{i}^{(m\pm\frac{1}{2})},
		\end{split} 
	\end{gather}
	and
	\begin{gather}
		\hat{I}^d_j\rvert_{(m\pm\frac{1}{2})} = - \frac{D^*_{m\pm\frac{1}{2}}}{\Delta x_j} \Delta_j (\rho e),
	\end{gather}     
	where $\hat{A_{\rho}}^{d}_j\rvert_{(m\pm{1}/{2})}$, $\hat{A_{\rho u}}^d_{ij}\rvert_{(m\pm{1}/{2})}$, and $\hat{A_{E}}^{d}_j\rvert_{(m\pm{1}/{2})} =\hat{K}^d_j\rvert_{(m\pm{1}/{2})} + \hat{I}^d_j\rvert_{(m\pm{1}/{2})}$ are the localized LF-type total discrete dissipative fluxes for the artificial terms in the mass, momentum, and energy equations [Eqs. \eqref{eq:art_mass}-\eqref{eq:art_energy}], respectively.

	

	%

	\section{Results\label{sec:results}}
	
	In this section, the proposed artificial-viscosity method is used to simulate a variety of test cases, involving (a) a classical one-dimensional shock-tube case, to assess the accuracy of this method in capturing the discontinuities and to illustrate the effect of the new sensor in Eq. \eqref{eq:new_sensor}, (b) an LES of compressible turbulent flow, to assess the low-dissipative nature with and without the use of the new sensor, and robustness in simulating compressible turbulent flows, (c) a shock-vortex interaction simulation, to assess the capability and accuracy of the method for simulating a more resolved DNS-type calculations, and (d) a drop advection simulation, to illustrate the advantages of the use of AMD over ATD for two-phase flows.
	
	The proposed method is implemented in the low-dissipative CTR-DIs3D solver \citep{jain2020conservative,jain2022kinetic}, which uses a second-order central scheme and a fourth-order Runge-Kutta scheme for spatial and temporal discretizations, respectively. 
	

	\subsection{Modified Sod test case}
	
	\begin{figure}
		\includegraphics[width=\textwidth]{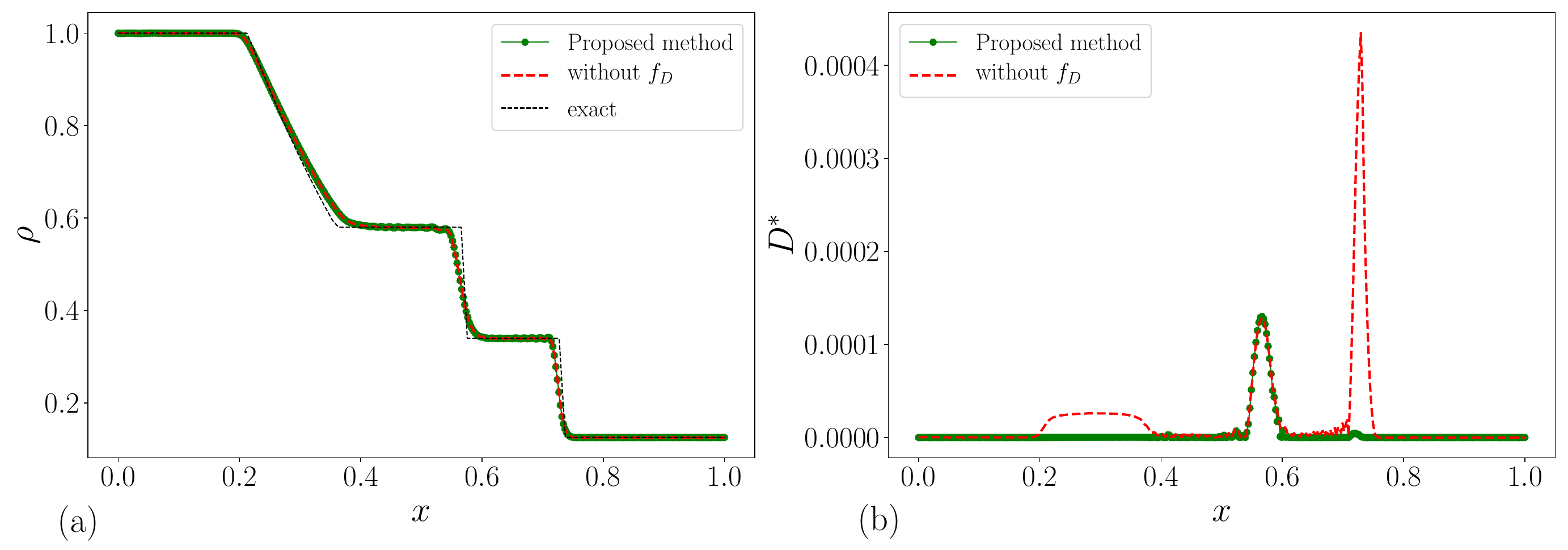}
		\caption{Modified version of the Sod shock-tube test case, simulated with the new sensor in Eq. \eqref{eq:new_sensor} (proposed method) and without the sensor (without $f_D$), showing: (a) density, $\rho$, and (b) artificial mass diffusivity, $D^*$.}
		\label{fig:mod_sod}
	\end{figure}
	The modified Sod-shock tube is a classic one-dimensional test case, used to assess the accuracy of shock and contact discontinuity-capturing methods, that was originally proposed by Ref. \cite{sod1978survey}. Here, a modified version of the test case is used, because an entropy-consistent scheme is needed to avoid the entropy-violating jump that would otherwise form in the expansion region of this modified Sod shock-tube case \citep{chandrashekar2013kinetic}. The initial setup consists of a Riemann problem with the left state $(\rho , u , p)=(1.0,0.75,1.0)$ and the right state $(\rho,u, p)=(0.125,0.0,0.1)$, and the discontinuity located at $x=0.3$. Note that a sharp discontinuity was not used at the initial time, instead the initial discontinuity was smoothed to $1-2$ grid cells thick. Starting with a sharp initial discontinuity can introduce locally unnecessarily large artificial fluid properties, which can impose a severe Courant-Friedrich-Lewy (CFL) restriction, which is a well known issue for LAD methods \citep{kawai2010assessment}. On the other hand, starting with a smooth discontinuity will not add large artificial fluid properties at time $t=0$, thereby minimizing the issue of severe CFL restriction. 
	The number of grid points is chosen to be $N=400$, and the results are presented at the final time of $t=0.2$ in Figure \ref{fig:mod_sod} along with the analytical (exact) solution. 
	
	The simulation results in Figure \ref{fig:mod_sod} show that the shock and contact discontinuities are captured accurately with the proposed LAD method, and that the method does not suffer from the formation of entropy-violating jump in the expansion region. The density, $\rho$, and artificial mass diffusivity, $D^*$, are plotted in Figure \ref{fig:mod_sod}(a) and \ref{fig:mod_sod}(b), respectively, at time $t=0.2$, with and without the newly proposed sensor, $f_D$. 
	Figure \ref{fig:mod_sod}(b) shows that the AMD ($D^*$) is active in the regions of shock, contact discontinuity, and expansion fan when the proposed $f_D$ sensor is not being used. However, the 
	use of the $f_D$ sensor localizes $D^*$ mostly to regions around the contact discontinuity, without introducing any more oscillations in the solution in the regions of shock (see Figure \ref{fig:mod_sod}(a)). Non-zero values of $D^*$ around shocks would unnecessarily make the method more dissipative because the ABV ($\beta^*$) is already active in this region to resolve the shock. Therefore, the use of the new switching sensor $f_D$ makes the method less dissipative without affecting the accuracy of the solution.
	
	\subsubsection{Sensitivity to model coefficients \label{sec:sensitivity1}}
	
\begin{figure}
		\includegraphics[width=\textwidth]{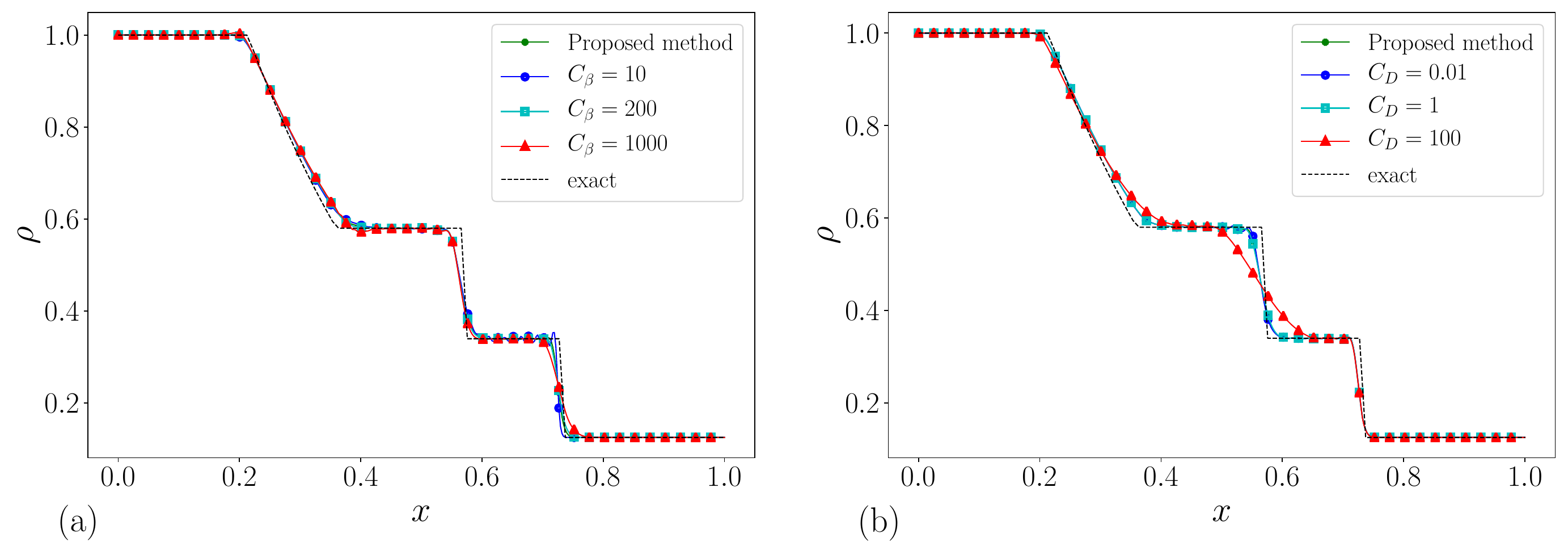}
		\caption{Modified version of the Sod shock-tube test case, showing density, for: (a) $C_{\beta}=$ 10, 100 (proposed method), 200, and 1000, and (b) $C_{D}=$ 0.01, 0.5 (proposed method), 1, and 100.}
		\label{fig:mod_sod_sensitivity}
	\end{figure}

	As mentioned in Section \ref{sec:coefficients}, the coefficient values are chosen to be $C_{D}=0.5$ and $C_{\beta}=100$ throughout this work with the second-order central scheme. It was claimed that the simulations are robust to these values and that the accuracy of the solution will not be significantly affected as long as the coefficient values are not changed by an order of magnitude. To illustrate this, the modified Sod-shock tube simulation is repeated for different values of the model coefficients in Figure \ref{fig:mod_sod_sensitivity}. 
	
	Figure \ref{fig:mod_sod_sensitivity} shows the solution density field for different values of model coefficients, $C_{\beta}$ and $C_D$ in ABV and AMD, respectively. Small oscillations can be seen around the shock for $C_{\beta}=10$ (10 times smaller than the proposed value), and the thickness of the shock is slightly increased for $C_{\beta}=1000$ (10 times higher than the proposed value). The contact discontinuity appears to be overly smeared for $C_D=100$ (200 times higher than the proposed value), but for smaller values of $C_D$, the contact discontinuity is intact and is not affected. 
	Overall, it can be seen that the solution is quite robust to the choices of both $C_{\beta}$ and $C_D$, and the accuracy of the solution is only affected when these model coefficients are varied by more than an order of magnitude away from the proposed values.
	
	\subsection{Decaying homogeneous isotropic turbulence with shocklets \label{sec:decaying_HIT}}
	
	In this section, the non-dissipative nature, accuracy, and robustness of the proposed artificial-viscosity method are assessed for the LES of compressible turbulent flows with shocklets. Here, a decaying homogeneous isotropic turbulence (HIT) is simulated at a high enough Mach number that shocklets are generated in the flow \citep{lee1991eddy}. In the incompressible limit of this flow, Agrawal et al. \cite{agrawal2022non} recently showed that the LES with dynamic Smagorinsky model accurately predicts the decay rate of the kinetic energy. For this reason, the dynamic Smagorinsky model is used in this work as well. 
	The initial Taylor-scale Reynolds number of the flow is $Re_{\lambda,o}=100$, and the initial turbulent Mach number is $M_{t,o}=0.6$. The initial conditions for this simulation are generated following the procedure described in Ref. \cite{johnsen2010assessment}. The Prandtl number is chosen to be $Pr =0.7$, and the material properties of the fluid are chosen to be $\gamma=1.4$ (specific heat ratio) and $R=1$ (specific gas constant) in the ideal gas law. 
	The domain is a triply periodic cube with dimensions $[0,2\pi]$.
	Here, a coarse resolution of $64^3$ grid points is used, and hence, this is a good test to assess the amount of numerical dissipation added by the shock--and contact discontinuity-capturing method. 

\begin{figure}
		\centering
		\includegraphics[width=\textwidth]{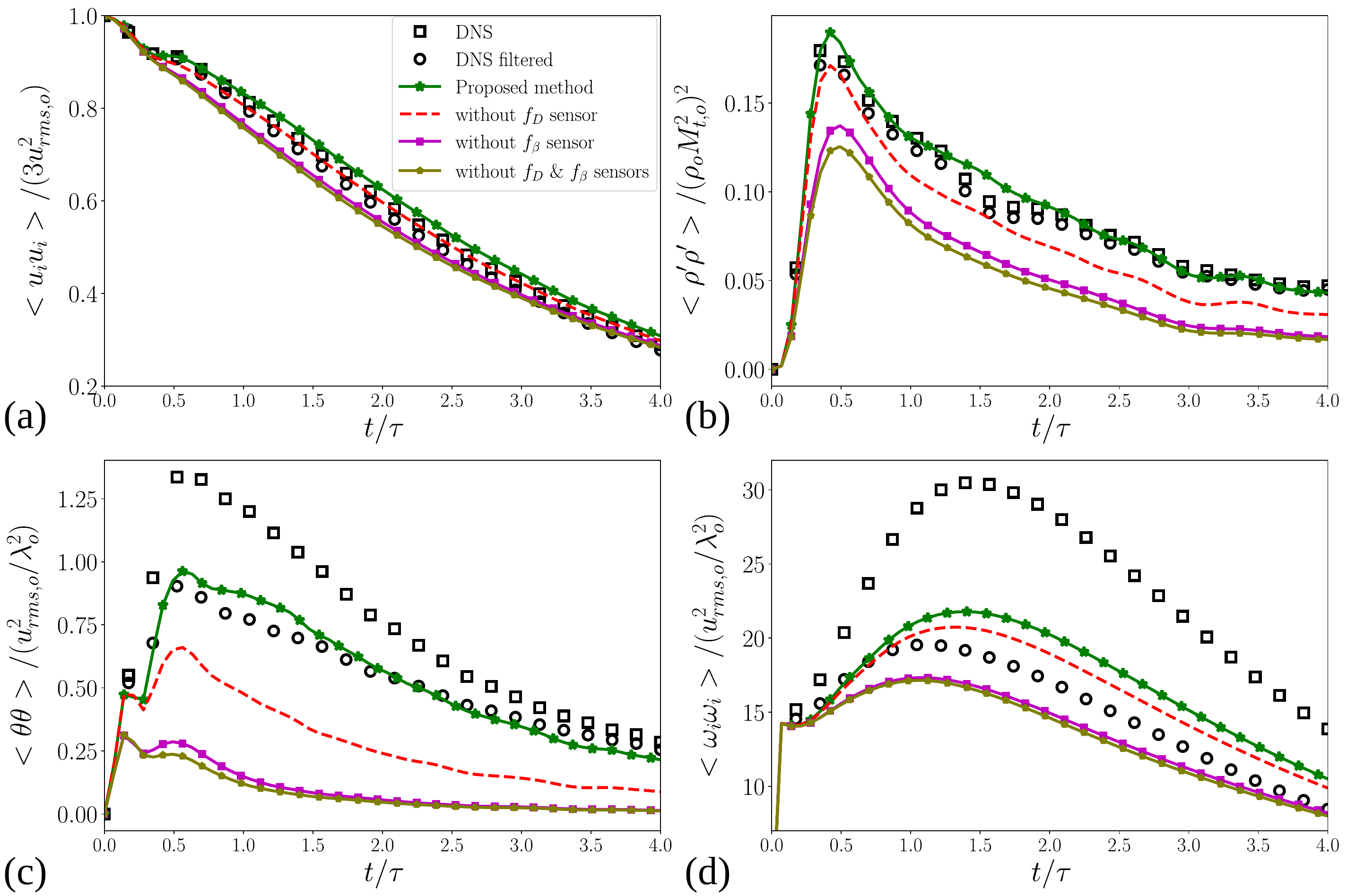}
		\caption{Simulation of decaying HIT with shocklets, with both the $f_D$ and $f_{\beta}$ sensors (proposed method), without the $f_D$ sensor (similar to the Ref. \cite{kawai2010assessment} formulation), without $f_{\beta}$ sensor, and without both the the $f_D$ and $f_{\beta}$ sensors (similar to the Ref. \cite{mani2009suitability} formulation), and its comparison with the DNS and filtered DNS. The plotted non-dimensional quantities: (a) mean square velocity, (b) density variance, (c) dilatational variance, and (d) vorticity variance.}
		\label{fig:decay_HIT}
	\end{figure}

	Figure \ref{fig:decay_HIT} shows the results from the simulation (a) with $f_D$ and $f_{\beta}$ (the proposed method), (b) with $f_D$ and without $f_{\beta}$, (c) without $f_D$ and with $f_{\beta}$ (similar to the formulation proposed in Ref. \cite{kawai2010assessment}, but note that here a modified form of the Ducros sensor is used instead of the original Ducros sensor that was used in Ref. \cite{kawai2010assessment}), and (d) without $f_D$ and without $f_{\beta}$ (similar to the formulation proposed in Ref. \cite{mani2009suitability}). To compare the results, a direct numerical simulation (DNS) of the same test case is performed on a $256^3$ mesh; the results are also filtered, using a Gaussian filter, onto a $64^3$ mesh and are shown in Figure \ref{fig:decay_HIT}. In Figure \ref{fig:decay_HIT} no subgrid model (ASV or eddy-viscosity model) is used for unresolved eddies in all the simulations, and hence, the vorticity variance is overpredicted. The effect of the use of a subgrid model for unresolved eddies is explored in Section \ref{sec:sgs_results}.
	
	Figure \ref{fig:decay_HIT} shows that the proposed method that uses both $f_D$ and $f_{\beta}$ sensors is the least dissipative of all the formulations. If the new $f_D$ sensor in Eq. \eqref{eq:new_sensor} is not used, the dilatational motions, the density fluctuations, and vortical motions are all significantly damped, with a greater effect on the dilatational motions and the density fluctuations. Similarly, if the modified Ducros sensor in Eq. \eqref{eq:Ducros_sensor} is not used, all the quantities are damped. Finally, if both $f_D$ and $f_{\beta}$ sensors are not used, the formulation is highly dissipative, making the simulations inaccurate. 
	
	Note that this observation is for the current choice of values for the model coefficients $C_{\beta}$ and $C_D$ (Section \ref{sec:coefficients}) used in this work, which were chosen such that the stronger shocks and contact discontinuities can be accurately captured with the current choice of the numerical scheme (see Section \ref{sec:sensitivity1}). If one chooses the values of $C_{\beta}$ and $C_D$ significantly lower than the proposed values, it is possible to perform a low dissipative simulation of decaying HIT even without the use of sensors. But these values of coefficients then will only work for turbulent flows with weak compressibility and will not be suitable for the simulation of stronger shocks and contact discontinuities. 
	Therefore, it is instead preferable to use the sensors to appropriately localize the dissipation than tuning the model coefficients for each problem.
	
\begin{figure}
		\centering
		\includegraphics[width=\textwidth]{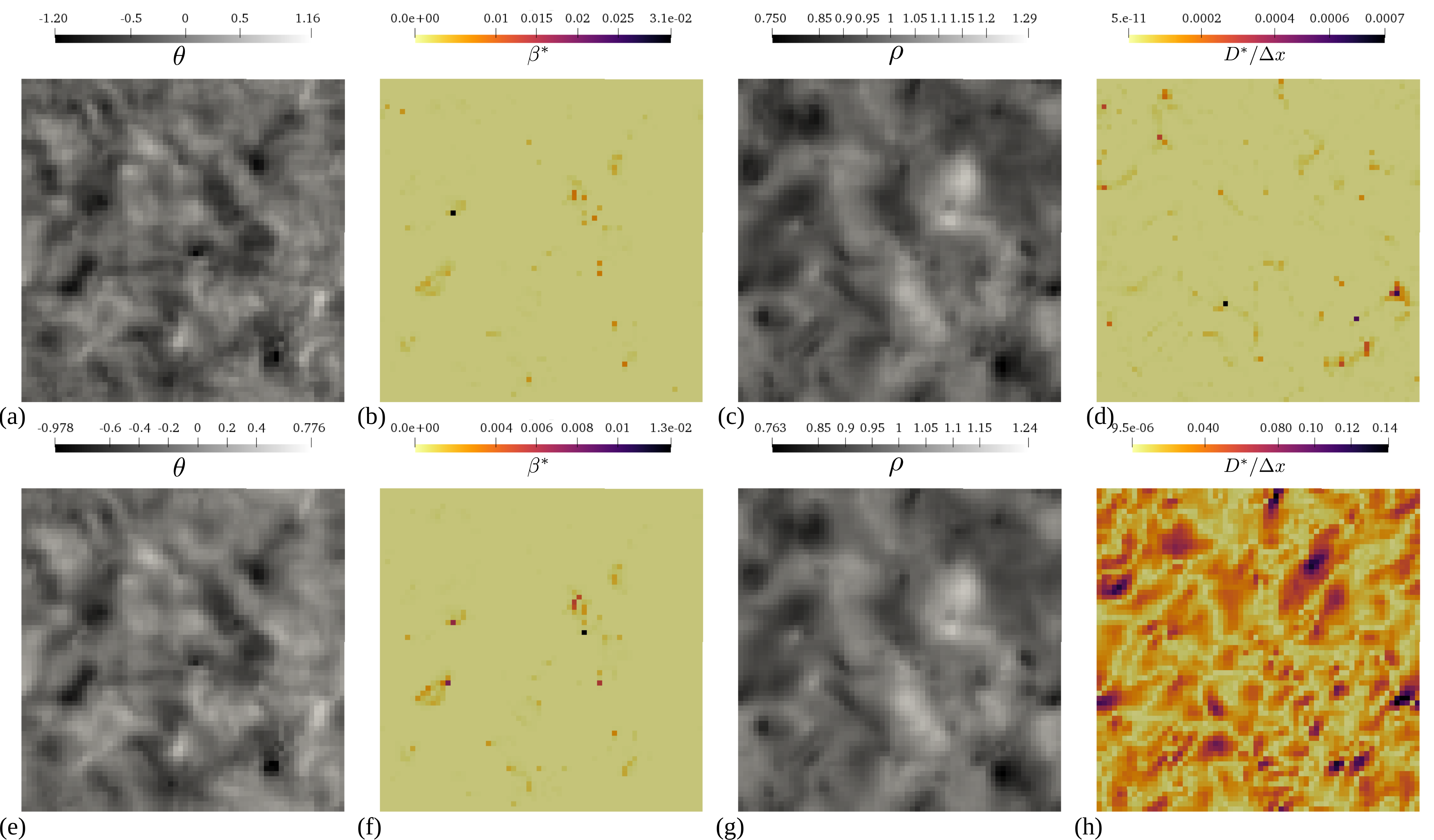}
		\caption{Two-dimensional slices from the simulations of decaying HIT with shocklets, showing (a)-(d) dilatation, ABV, density, and AMD with the $f_D$ sensor, and (e)-(h) dilatation, ABV, density, and AMD without the $f_D$ sensor.}
		\label{fig:2D_slices}
	\end{figure}

	The two-dimensional slices from the simulation, with and without the $f_D$ sensor, are plotted in Figure \ref{fig:2D_slices}. With the $f_D$ sensor active, the AMD can be seen to be significantly reduced by almost more than 2 orders of magnitude without spuriously affecting the scales of density fluctuations and dilatational motions. In fact, without the $f_D$ sensor, the dilatational field and the density field are seen to be damped slightly (see the range in the color bar). The ABV is seen to be active in the regions of high dilatation, as expected, and without the $f_D$ sensor, the ABV is increased by a small amount to compensate for the reduced AMD. But the overall structure of the regions where ABV is active is still the same, with and without the $f_D$ sensor.
	
	In summary, the current simulation results show that the use of $f_D$ and $f_{\beta}$ sensors are necessary to recover the correct behavior of kinetic energy, and variances of dilatation, density, and vorticity.  
	The simulations are the least accurate without the use of $f_D$ and $f_{\beta}$ sensors. Using these $f_D$ and $f_{\beta}$ sensors individually improves the results by making the method less dissipative, and using both sensors gives the best results. 
	Therefore, the proposed method along with the sensors results in a robust, accurate, and low-dissipative method for capturing shocks and contact discontinuities for LES of compressible turbulent flows.


	
	
	
	\subsubsection{Effect of an eddy-viscosity model \label{sec:sgs_results}}
	
\begin{figure}
		\centering
		\includegraphics[width=\textwidth]{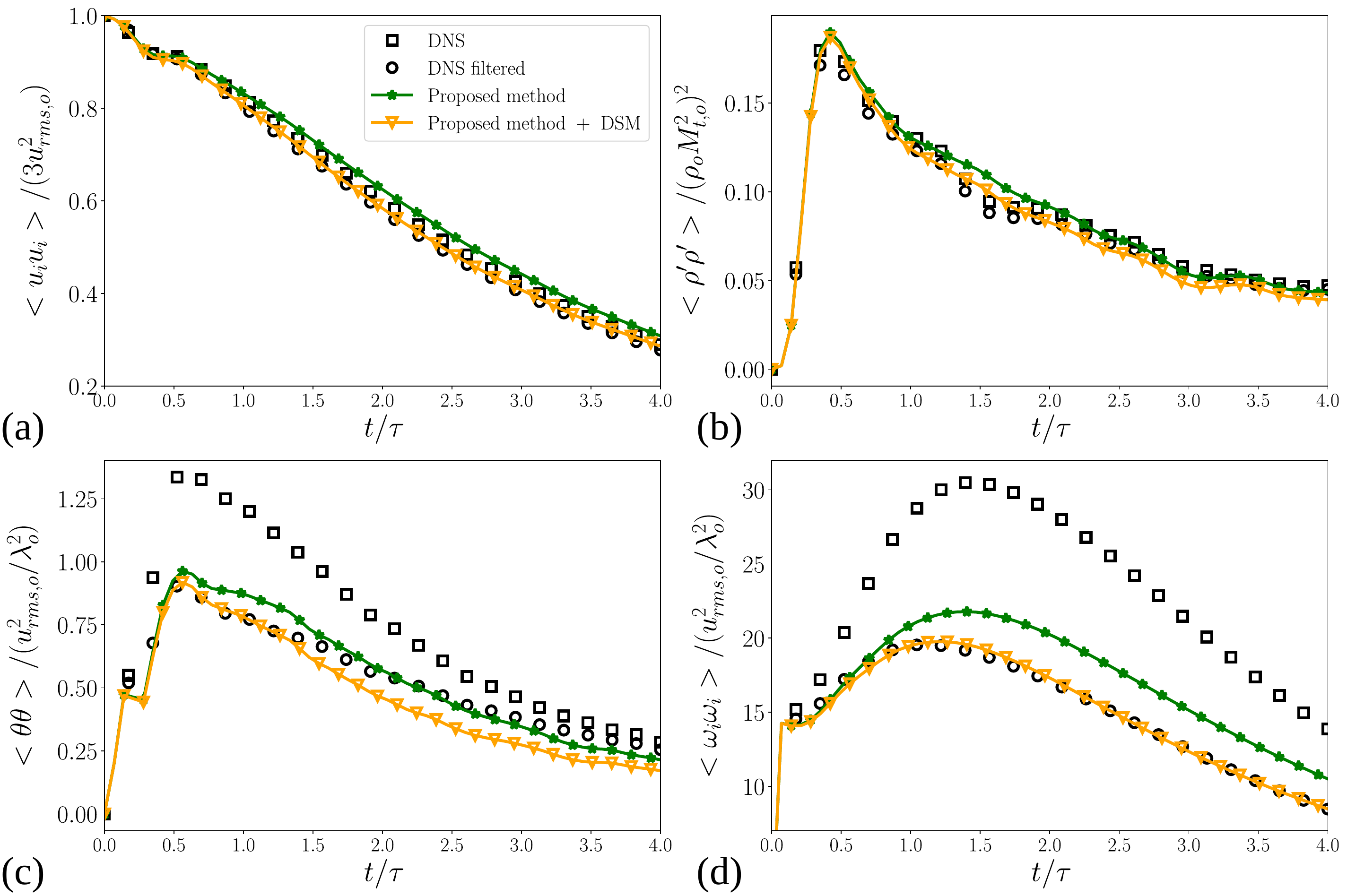}
		\caption{Simulation of decaying HIT with shocklets, with both the $f_D$ and $f_{\beta}$ sensors, and without DSM (proposed method) and with DSM (proposed method + DSM), showing the non-dimensional quantities: (a) Mean square velocity, (b) density variance, (c) dilatational variance, and (d) vorticity variance.}
		\label{fig:decay_HIT_sgs}
	\end{figure}

	Since this is an LES of a turbulent flow, a subgrid model (either an ASV or an eddy-viscosity model) is required for unresolved eddies. So far, no model was used in Figure \ref{fig:decay_HIT} to isolate only the effect of ABV and AMD and to illustrate the effect of the use of the sensors $f_{\beta}$ and $f_D$. Here, the LES of decaying HIT is repeated with the use of a dynamic Smagorinsky model described in Section \ref{sec:dsm} to see the effect of this model. 
	Figure \ref{fig:decay_HIT_sgs} shows the simulation results for the proposed method, that uses both $f_D$ and $f_{\beta}$ sensors, with and without the dynamic Smagorinsky model (DSM).
	
	The use of the DSM has an important effect on kinetic energy and all the variances in Figure \ref{fig:decay_HIT_sgs}. Without the DSM, all the quantities are overpredicted, but with the use of DSM, the simulation is most accurate with all the quantities following very closely with the predicted filtered DNS data. Particularly, the accuracy of the evolution of kinetic energy and vorticity variance is greatly improved with the use of DSM. It is important to also appreciate here that this is only possible because of the low-dissipative nature of the proposed LAD method, with the new sensors that appropriately localize the dissipation to the regions where it's needed. Without the use of $f_D$ and $f_{\beta}$ sensors, the formulation was very dissipative, as seen in Figure \ref{fig:decay_HIT}, and with the addition of DSM, it would only become more dissipative, significantly affecting the accuracy of the simulation.

\begin{figure}
		\centering
		\includegraphics[width=\textwidth]{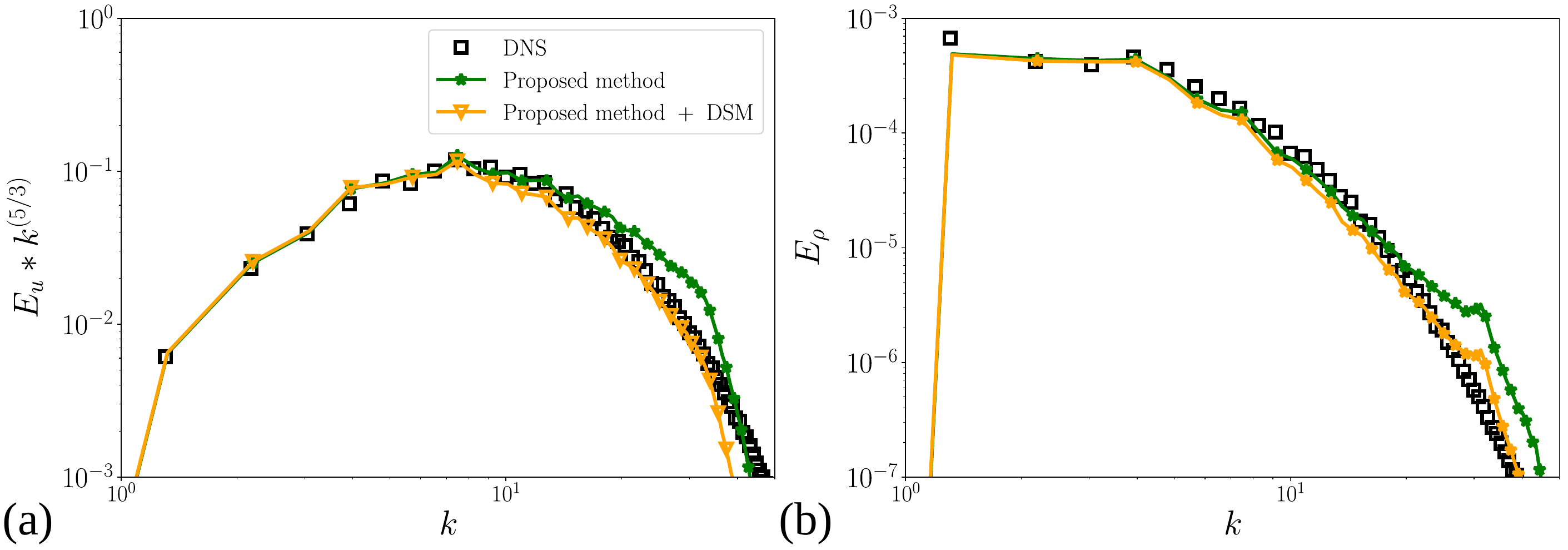}
		\caption{Simulation of decaying HIT with shocklets, with both the $f_D$ and $f_{\beta}$ sensors, and without DSM (proposed method) and with DSM (proposed method + DSM), showing the spectrum of: (a) kinetic energy and (b) density variance.}
		\label{fig:spectrum}
	\end{figure}

	In addition to the integrated quantities, we plot the spectrum of kinetic energy and density variance in Figure \ref{fig:spectrum}, with and without DSM, which gives a more detailed information on how the models act at different scales. It shows that the use of DSM along with the proposed LAD approach clearly improves the result by bringing the spectra closer to the DNS spectra, except with deviations at small scales which are to be expected in a LES calculation.
	
	\subsubsection{Sensitivity to model coefficients \label{sec:sensitivity2}}
	
\begin{figure}
		\centering
		\includegraphics[width=\textwidth]{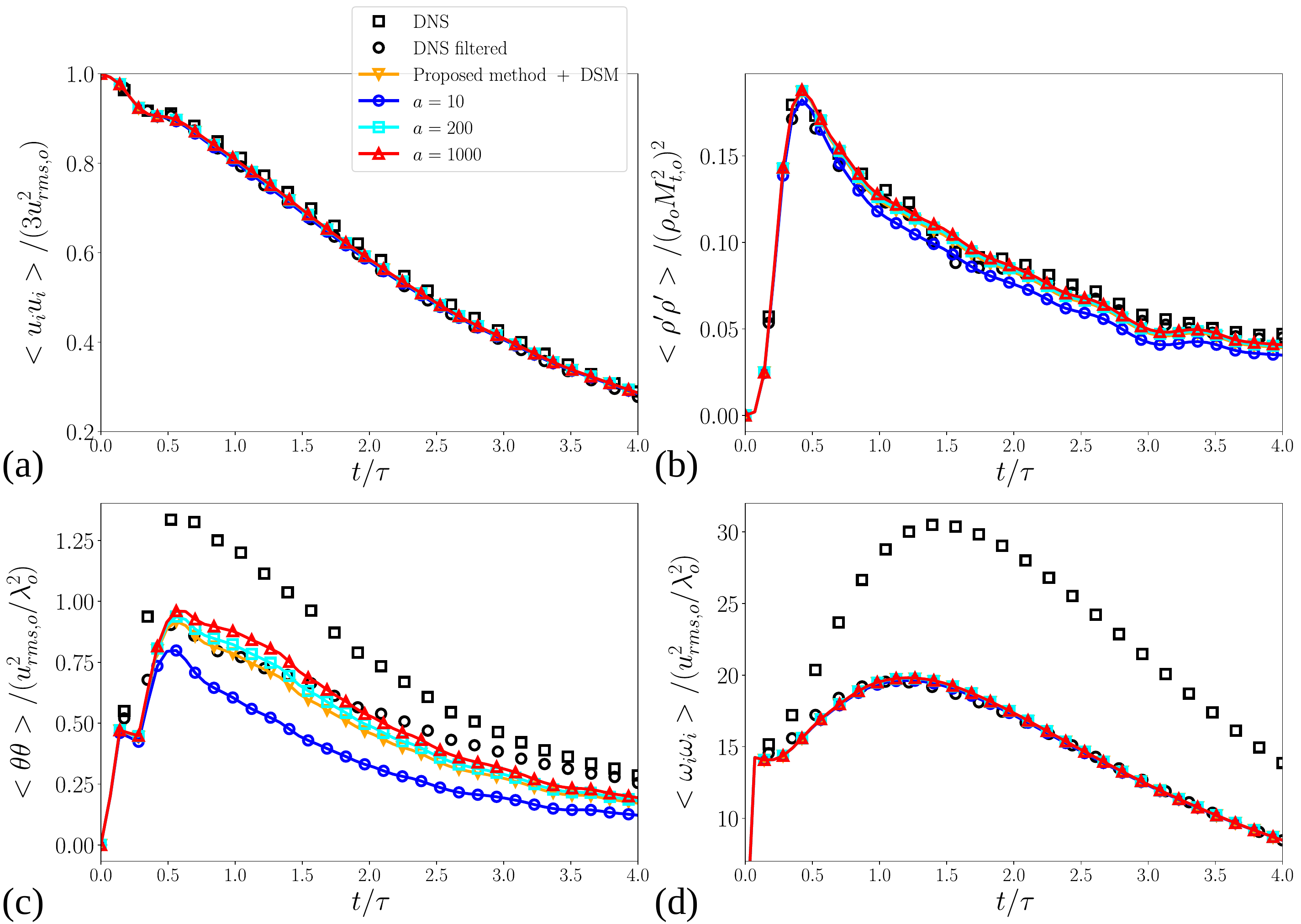}
		\caption{Simulation of decaying HIT with shocklets with $f_D$, $f_{\beta}$, and DSM active, for $a=$10, 100 (proposed method + DSM), 200, and 1000, showing the non-dimensional quantities: (a) mean square velocity, (b) density variance, (c) dilatational variance, and (d) vorticity variance.}
		\label{fig:decay_HIT_sgs_sensitivity}
	\end{figure}

	The sensitivity to the model coefficients $C_D$ and $C_{\beta}$ was already tested in Section \ref{sec:sensitivity1}. Here, the sensitivity to the model coefficient $a$ in the $f_D$ and $f_{\beta}$ sensors in Eqs. \ref{eq:Ducros_sensor}, \eqref{eq:new_sensor} is tested. Here, $a$ is added in the sensors to make the sensor stronger and to further localize the regions where ABV and AMD/ATD are active. To illustrate this, the decaying HIT simulation is repeated for different values of $a$ in Figure \ref{fig:decay_HIT_sgs_sensitivity}.
	
	Figure \ref{fig:decay_HIT_sgs_sensitivity} shows the results from the decaying HIT simulation, with $f_D$, $f_{\beta}$, and DSM active, for $a=$10, 100 (proposed method + DSM), 200, and 1000. Clearly, varying $a$ only affects the dilatational motions and the density fluctuations, and the effect is relatively small. The biggest improvement is seen when the value of $a$ is increased from 10 to 100 and the solution becomes less dissipative and is closer to the predicted filtered DNS data, and beyond this it saturates, a further increase in the value of $a$ only has a minimal effect on the solution. Hence, a value of $a=100$ is chosen in this work. 
	
	\subsection{Shock-vortex interaction}
	
	In this section, a shock-vortex interaction is simulated using the proposed artificial-viscosity method. Section \ref{sec:decaying_HIT} demonstrated that the proposed method is low-dissipative and is suitable for LES of compressible turbulent flows. This section, in contrast, will assess the accuracy and suitability of the present method for more resolved simulations and DNS of compressible turbulent flows. 
	
\begin{figure}
		\centering
		\includegraphics[width=0.3\textwidth]{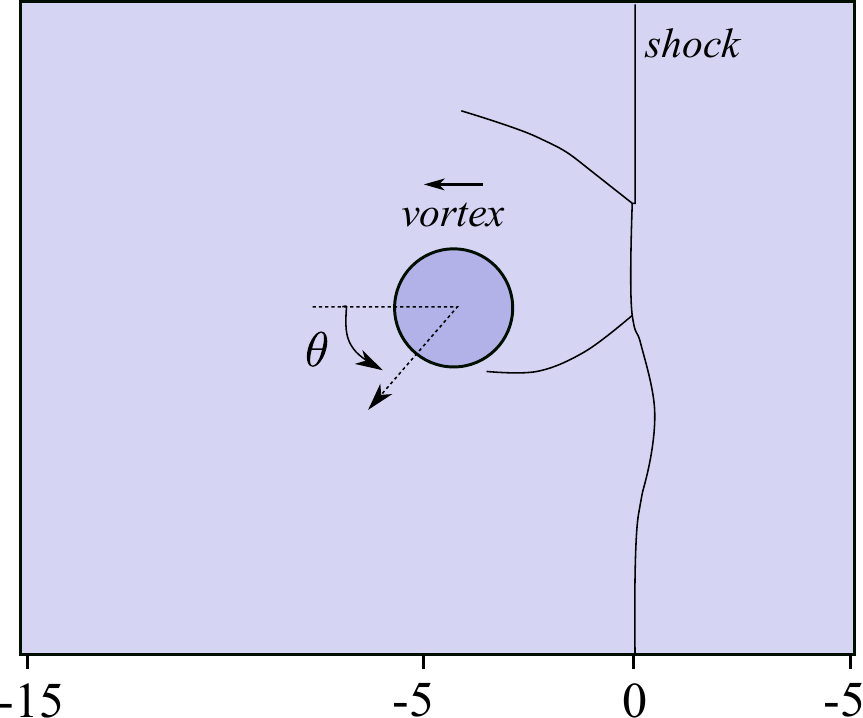}
		\caption{Schematic of the problem setup for the shock-vortex interaction simulation, showing the moving vortex and the deformed stationary shock after the vortex has passed through the shock.}
		\label{fig:schematic}
	\end{figure}

\begin{figure}
		\centering
		\includegraphics[width=\textwidth]{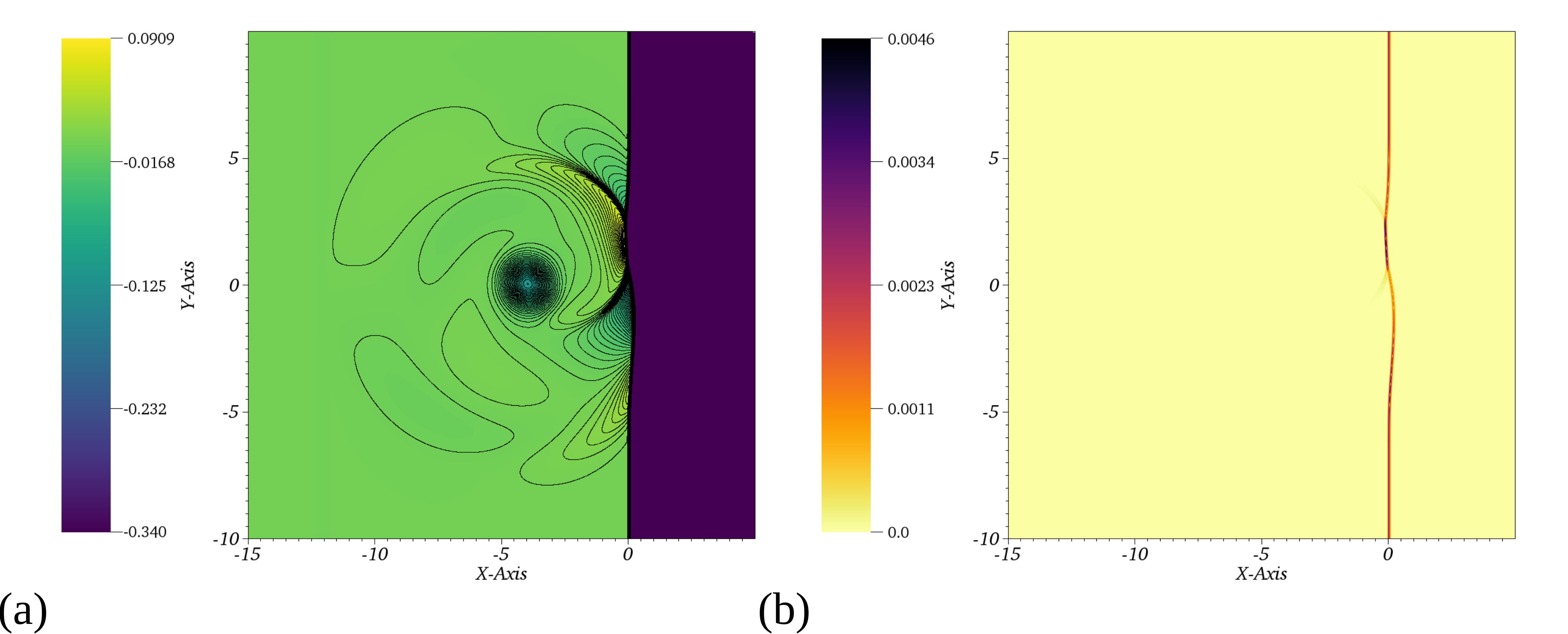}
		\caption{Simulation of a shock-vortex interaction, showing: (a) the sound pressure field, $\Delta p = (p-p_{\infty})/p_{\infty}$ and (b) the artificial bulk viscosity, $\beta^*$, at $t=6$.}
		\label{fig:shock_vortex}
	\end{figure}

	This test case is taken from the work of Ref. \cite{inoue1999sound}, Ref. \cite{zhang2005multistage}, and Ref. \cite{chatterjee2008multiple} and has also been used to evaluate the shock-capturing capability by Ref. \cite{subramaniam2019high} and Ref. \cite{haga2019robust}. The initial setup of this case consists of a $M=1.2$ stationary shock located at $x=0$ and an isentropic vortex of strength $M_v=0.25$, initially located upstream of the shock at $x=2$. The domain extent is $[-30,10]\times[-20,20]$.
	The initial vortex field is given by
	\begin{gather}
		u_{\theta}(r) = M_v r \exp{\left(\frac{1 - r^2}{2}\right)}, \\
		u_r(r) = 0,\\
		p(r) = \frac{1}{\gamma} \left[ 1 - \left(\frac{\gamma - 1}{2}\right) M_v^2 \exp{(1 - r^2)} \right]^{\frac{\gamma}{\gamma - 1}},
	\end{gather}
	and
	\begin{gather}
		\rho(r) = \left[ 1 - \left(\frac{\gamma - 1}{2}\right) M_v^2 \exp{(1 - r^2)} \right]^{\frac{1}{\gamma - 1}},
	\end{gather}
	where $u_{\theta}$ is the angular velocity, $u_r$ is the radial velocity, $r$ is the radial distance from the center of the vortex, and $\gamma$ is the ratio of specific heats. The schematic of the setup after the vortex has passed through the shock is shown in figure \ref{fig:schematic}.
	The sound pressure field at the time of $t=6$ is shown in Figure \ref{fig:shock_vortex} along with the ABV, $\beta^*$. Here, in this simulation, AMD is also active along with the $f_D$ sensor. At time $t=6$, the vortex has passed through the shock and deforms the shock surface. The results show that the $f_{\beta}$ sensor has successfully localized ABV only to those regions around the shock, while maintaining the accuracy in capturing the shock without the need for separately tuning the coefficients or the need to turn off AMD in this case. 
	
\begin{figure}
		\centering
		\includegraphics[width=\textwidth]{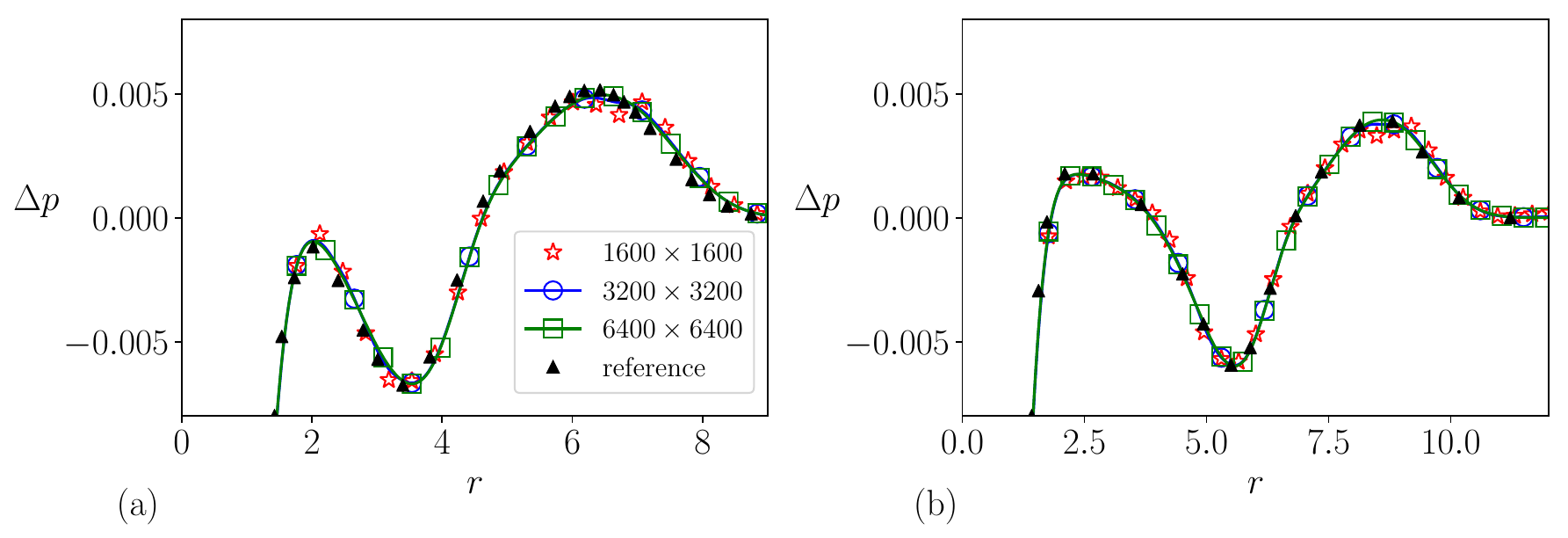}
		\caption{Simulation of a shock-vortex interaction, showing the sound pressure field, $\Delta p = (p-p_{\infty})/p_{\infty}$ along the radial direction $r$ from the center of the vortex for three grid resolutions, along with the reference solution from Ref. \cite{inoue1999sound}, and at (a) $t=6$ (b) $t=8$.}
		\label{fig:shock_vortex_radial}
	\end{figure}

	The sound pressure is also plotted in Figure \ref{fig:shock_vortex_radial} along the radial direction $r$ from the vortex center for a fixed value of $\theta=-45^{\circ}$ at two different times ($t=6, 8$) and for three grid resolutions against the reference solution from Ref. \cite{inoue1999sound}. The plots show that the simulations are grid converged and it matches well with the reference solution. 
	
	\subsection{Droplet advection}
	
	In this section, a droplet advection is simulated to assess the suitability of the proposed artificial-viscosity method for two-phase flows. The LAD formulation was extended to two-phase flows in Section \ref{sec:two_phase} for a five-equation model \citep{jain2020conservative} or a four-equation model \citep{jain2023assessment} that can be used with a central-difference scheme.
	
	The five-equation model doesn't assume thermal equilibrium, and admits two temperatures for two phases. Therefore, ATD cannot be directly used with a five-equation model, which would lead to the violation of interface-equilibrium condition (IEC). However, AMD is constructed in such a way that it satisfies IEC with a five-equation model (see Appendix A). 
	
	Here, the use of AMD and ATD with a five-equation model for a simple one-dimensional advection of the drop is assessed. The domain length is 5 units and has periodic boundary conditions on both sides. A drop of radius 0.5 units is initially placed at the center of the domain at $x=2.5$.
	The initial velocity is prescribed to be $u=2.5$ and the initial pressure is $p=1$. 
	Since both the velocity and pressure is uniform at the initial time, the pressure and velocity has to remain uniform for all times (definition of IEC). 
	
\begin{figure}
		\centering
		\includegraphics[width=\textwidth]{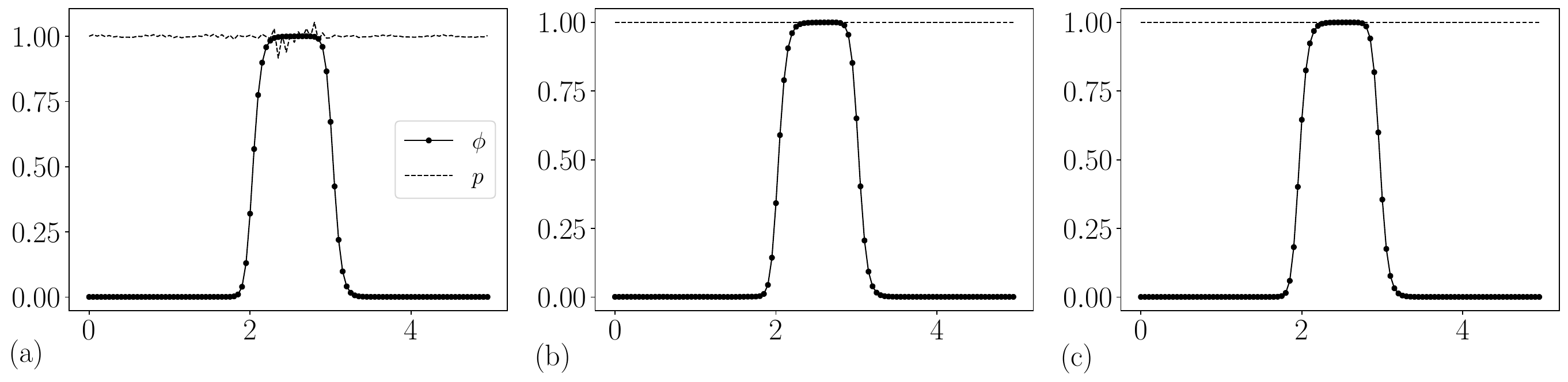}
		\caption{Simulation of a one-dimensional drop advection, showing the volume fraction and pressure fields at time $t=2$, with: (a) ATD, (b) no ATD or AMD, (c) AMD.}
		\label{fig:drop_advection}
	\end{figure}

	Here, the material interface is the only discontinuity in this problem, and since the phase-field model \citep{jain2022accurate} is already acting to capture the material interface, the artificial viscosities (ABV and AMD) should not have any effect on this problem. Figure \ref{fig:drop_advection} shows the pressure and volume fraction at the final time of $t=2$ with the use of ATD, AMD, and without ATD or AMD. 
	When no ATD or AMD is used, the pressure remains uniform as expected. 
	Since ATD violates IEC, spurious oscillations in the pressure field can be seen at the interface location when ATD is used. But when AMD is used, the pressure remains uniform because AMD satisfies IEC, similar to when no ATD or AMD is used. Hence, it is recommended to use AMD for capturing contact discontinuities in two-phase flows.
	
	\section{Summary and conclusions\label{sec:conclusions}} 
	
	
	
	
	In this work, we propose a novel, entropy-consistent, and stable localized artificial-viscosity/diffusivity (LAD) method for capturing shocks and contact discontinuities in compressible flows. 
	The artificial-viscosity methods have many advantages over other discontinuity capturing methods due to its simplicity and the ability to be used with central-difference schemes, which is beneficial for the simulation of turbulent flows. But the main challenge lies in (a) the need for appropriate localization of the dissipation to regions of discontinuity where it is required, (b) the need for sophisticated filtering operation for the artificial fluid properties to stabilize the method, which will make it difficult to extend the formulation for unstructured grids, and (c) the need for tuning the coefficients and to turn on-and-off the artificial fluid properties depending on the problem. The proposed method in this work overcomes most of these challenges.
	
	Moreover, all existing LAD methods are used with a high-order central scheme, and it is not clear how these methods perform with low-order schemes. In this work, a second-order central scheme is chosen because of its low cost; low aliasing error; easy implementation, boundary treatment, and extension to unstructured grids; and improved stability. Guidelines on how to choose the parameters in the method are provided.
	
	Furthermore, in artificial mass/thermal diffusivity (AMD/ATD), either the density or internal energy was used previously as the indicator function to detect contact discontinuities. But the main issue with this sensor is that it not only activates in the regions of contact discontinuity, but also activates in the regions of shock and vertical motions. This would be a problem because the artificial bulk viscosity (ABV) is already active in the regions of shock and a subgrid model is already active for unresolved eddies, and therefore, adding artificial mass/thermal diffusivity here will be unnecessarily dissipative. To prevent this from happening, we propose a sensor, "like" a Ducros sensor, that can effectively distinguish contact discontinuities from shocks and vortical motions, and turns on the AMD/ATD only in the regions of contact discontinuities. In ABV, we also use a stronger modified Ducros sensor instead of the original sensor, which further localizes ABV to only the regions of shock. 
	
	The use of these sensors in AMD/ATD and ABV will result in a LAD method that does not require tuning of the model coefficients, which is otherwise required, depending on the problem being solved.
	We show that the proposed method accurately captures shocks and contact discontinuities, without the need for problem-dependent tuning, for a range of problems, such as a one-dimensional Sod test case, decaying homogeneous isotropic turbulence with shocklets, and shock-vortex interaction.
	
	Using an analogy between the Lax-Friedrichs (LF) flux and the artificial-viscosity methods, a discrete LF-type flux formulation is presented for the proposed LAD method that satisfies discrete consistency conditions for kinetic energy and entropy. This LAD formulation is then coupled with the system of equations for compressible flows that are discretized using a robust central scheme. This results into a stable, low-dissipative, artificial-viscosity formulation which does not require filtering the solution and the artificial fluid properties that were previously required to obtain stable solutions.
	Therefore, the proposed method is suitable for LES and DNS of compressible turbulent flows with discontinuities in complex geometries. 
	
	An extension of the proposed method to capture shocks and contact discontinuities in compressible two-phase flows is also presented. It is shown that the proposed method satisfies the interface equilibrium condition, a crucial thermodynamic consistency condition for robust numerical simulations of compressible two-phase flows.





	

	\section*{Appendix A: Interface equilibrium condition}
	
	
	According to the definition of IEC, if velocity and pressure are initially uniform, they have to remain uniform at all times. AMD satisfies IEC, but ATD does not satisfy IEC. 
	Hence, it might be preferable to use AMD for two-phase flows because it satisfies IEC, irrespective of the choice of the model.
	
	To show this, let us consider only the LAD terms in Eqs. \eqref{eq:volumef}-\eqref{eq:energyf} and ignore other terms because they satisfy IEC \citep{jain2020conservative}, and make a one-dimensional assumption.
	
	\subsection{LAD with AMD and ABV}
	
	The simplified system of equations, with only AMD and ABV terms, can be written as
	\begin{equation}
		\frac{\partial \phi}{\partial t} = \frac{\partial}{\partial x} \left( D^* \frac{\partial \phi}{\partial x}\right),
		\label{eq:amd_vol}
	\end{equation}
	\begin{equation}
		\frac{\partial \rho}{\partial t} = \frac{\partial}{\partial x} \left( D^* \frac{\partial \rho}{\partial x}\right),
		\label{eq:amd_rho}
	\end{equation}
	\begin{equation}
		\frac{\partial \rho u}{\partial t} = \frac{\partial}{\partial x} \left( D^* u \frac{\partial \rho}{\partial x}\right) + \frac{\partial}{\partial x} \left( \beta^* \frac{\partial u}{\partial x}\right),
		\label{eq:amd_mom}
	\end{equation}
	\begin{equation}
		\frac{\partial E}{\partial t} = \frac{\partial}{\partial x} \left( D^* k \frac{\partial \rho}{\partial x}\right)  + \frac{\partial}{\partial x} \left( D^* \frac{\partial \rho e}{\partial x}\right) + \frac{\partial}{\partial x} \left( \beta^* u \frac{\partial u}{\partial x}\right).
		\label{eq:amd_energy}
	\end{equation}
	Assuming $u=constant$ initially in Eq. \eqref{eq:amd_mom}, it can be rewritten as
	\begin{equation}
		\rho \frac{\partial u}{\partial t} + u\left[\frac{\partial \rho}{\partial t} = \frac{\partial}{\partial x} \left( D^* \frac{\partial \rho}{\partial x}\right)\right],
	\end{equation}
	where the relation in within $[\cdot]$ is $0$ because of Eq. \eqref{eq:amd_rho}. Hence, $\partial u/\partial t=0$ and $u$ remains constant.
	Now, taking a dot product of velocity with the momentum equation in Eq. \eqref{eq:amd_mom}, the kinetic energy can be derived as
	\begin{equation}
		\frac{\partial \rho k}{\partial t} = \frac{\partial}{\partial x} \left( D^* k \frac{\partial \rho}{\partial x}\right) + u \frac{\partial}{\partial x} \left( \beta^* \frac{\partial u}{\partial x}\right).
		\label{eq:amd_ke}
	\end{equation}
	The internal energy equation can be obtained by subtracting the kinetic energy equation in Eq. \ref{eq:amd_ke} from the total energy equation in Eq. \eqref{eq:amd_energy} as
	\begin{equation}
		\frac{\partial \rho e}{\partial t} = \frac{\partial}{\partial x} \left( D^* \frac{\partial \rho e}{\partial x}\right) + \beta^*  \left(  \frac{\partial u}{\partial x}\right)^2,
		\label{eq:amd_ie}
	\end{equation}
	
	
	In a five-equation model, thermal equilibrium is not assumed, and each phase has its own temperature $T_l$, but an isobaric closure law is assumed to close the system of equations. Now, expressing $\rho e$ using the mixture rule, assuming ideal gas law for each phase, invoking isobaric law, internal can be written as
	\begin{equation}
		\rho e = \sum_l \rho_l e_l \phi_l = \sum_l \frac{\phi_l p_l}{\gamma_l - 1} = p \sum_l \frac{\phi_l}{\gamma_l - 1} = p \alpha,
		\label{eq:amd_1}
	\end{equation}
	where $\alpha = \sum_l {\phi_l}/({\gamma_l - 1})$. Using Eq. \eqref{eq:amd_1} and rewriting Eq. \eqref{eq:amd_ie} in terms of $p$, we get
	\begin{equation}
		\frac{\partial p \alpha}{\partial t} = \frac{\partial}{\partial x} \left( D^* \frac{\partial p \alpha}{\partial x}\right) + \beta^*  \left(  \frac{\partial u}{\partial x}\right)^2,
	\end{equation}
	Now, assuming $u=p=constants$ at initial time, we can rewrite this as
	\begin{equation}
		\alpha \frac{\partial p }{\partial t} + p \left[ \frac{\partial \alpha }{\partial t} = \frac{\partial}{\partial x} \left( D^* \frac{\partial \alpha}{\partial x}\right) \right],
	\end{equation}
	where the relation in within $[\cdot]$ is $0$ because it satisfies Eq. \eqref{eq:amd_vol}. Hence, $\partial p/\partial t=0$ and $p$ remains constant, and therefore IEC is satisfied with AMD.
	
	\subsection{LAD with ATD and ABV}
	
	Similarly, the simplified system of equations, with only ATD and ABV terms, can be written as
	\begin{equation}
		\frac{\partial \phi}{\partial t} = 0,
		\label{eq:atd_vol}
	\end{equation}
	\begin{equation}
		\frac{\partial \rho}{\partial t} = 0,
		\label{eq:atd_rho}
	\end{equation}
	\begin{equation}
		\frac{\partial \rho u}{\partial t} = \frac{\partial}{\partial x} \left( \beta^* \frac{\partial u}{\partial x}\right),
		\label{eq:atd_mom}
	\end{equation}
	\begin{equation}
		\frac{\partial E}{\partial t} = \frac{\partial}{\partial x} \left( D^* \frac{\partial \rho e}{\partial x}\right) + \frac{\partial}{\partial x} \left( \beta^* u \frac{\partial u}{\partial x}\right),
		\label{eq:atd_energy}
	\end{equation}
	\begin{equation}
		\frac{\partial \rho k}{\partial t} = u \frac{\partial}{\partial x} \left( \beta^* \frac{\partial u}{\partial x}\right),
		\label{eq:atd_ke}
	\end{equation}
	\begin{equation}
		\frac{\partial \rho e}{\partial t} = \frac{\partial}{\partial x} \left( D^* \frac{\partial \rho e}{\partial x}\right) + \beta^*  \left(  \frac{\partial u}{\partial x}\right)^2.
		\label{eq:atd_ie}
	\end{equation}
	Assuming $u=constant$ initially, from Eqs. \eqref{eq:atd_rho}, \eqref{eq:atd_mom}, it is easy to see that $\partial u/\partial t=0$, and therefore, $u$ remains constant.
	Now, rewriting Eq. \eqref{eq:atd_ie} in terms of $p$ and assuming $u=p=constants$ at initial time, we can rewrite it as
	\begin{equation}
		\alpha \frac{\partial p }{\partial t} + p \left[ \frac{\partial \alpha }{\partial t} = \frac{\partial}{\partial x} \left( D^* \frac{\partial \alpha}{\partial x}\right) \right].
	\end{equation}
	However, the relation in within $[\cdot]$ is not $0$ here, unlike in the case of AMD, because it doesn't satisfy Eq. \eqref{eq:atd_vol}. Hence, $\partial p/\partial t \ne 0$, and therefore, IEC is not satisfied with ATD.
	
	\section*{Acknowledgments} 
	
	S. S. J. gratefully acknowledges partial financial support from the Franklin P. and Caroline M. Johnson Graduate Fellowship and Boeing Co.. R.A. and P.M. acknowledge support from NASA's Transformational Tools and Technologies project grant, \#80NSSC20M0201.  
	S. S. J. thanks Tim Flint for providing helpful comments on this work and for helping with the exact solution of the Sod shock-tube test case, and acknowledges fruitful discussions with Henry Collis.

	\bibliographystyle{unsrt}
	\bibliography{shock}

\end{document}